\documentclass[12pt]{article}
\pdfoutput =1
\usepackage[left=18mm,top=20mm,right=18mm,bottom=15mm]{geometry}
\usepackage{graphicx}
\usepackage{color}
\usepackage{setspace}
\usepackage{float} 
\usepackage{amsmath}
\usepackage{slashed}
\usepackage{amsfonts}   
\usepackage{amssymb}
\usepackage{enumerate}
\usepackage{physics}
\usepackage{mathrsfs}
\usepackage{xcolor}
\usepackage{mdframed}
\definecolor{darkcyan}{rgb}{0.0, 0.55, 0.55}
\definecolor{darkspringgreen}{rgb}{0.09, 0.45, 0.27}
\definecolor{tropicalrainforest}{rgb}{0.0, 0.46, 0.37}
\definecolor{tealbright}{rgb}{0,128,128}
\definecolor{tealblue}{rgb}{30,108,135}
\definecolor{ao}{rgb}{0,0.5,0}
\definecolor{amethyst}{rgb}{0.6,0.4,0.8}
\usepackage[colorlinks]{hyperref}
\hypersetup{
  colorlinks,
  citecolor=blue,
  linkcolor=red,
  urlcolor=violet}
\usepackage{upgreek}
\usepackage{float}
\usepackage{subfloat}
\usepackage{tikz}
\usetikzlibrary{positioning}
\usepackage{authblk}

\title{Entanglement measures for non-conformal D-branes}
\author{Arindam Lala\thanks{arindam.physics1@gmail.com, arindam.lala@pucv.cl}}
\affil{\small Instituto de F\'{i}sica, \protect\\ Pontificia Universidad Cat\'{o}lica de Valpara\'{i}so,
\protect\\ Casilla 4059, Valpara\'{i}so, Chile}

\begin{document}
\date{}
\renewcommand{\a}{\alpha}
\renewcommand{\b}{\upbeta}
\renewcommand{\d}{\delta}

\maketitle
\begin{abstract}
We study various entanglement measures associated with certain non-conformal field theories. We consider
non-conformal D$p$-brane backgrounds, which are dual to these field theories, for our holographic analysis.
Restricting our interests in $p=1,2,4$, we explicitly compute properties of holographic entanglement entropy
and entanglement wedge cross section, $\text{E}_{W}$, corresponding to two parallel strip shaped boundary
subregions in these set ups. We study low and high temperature behaviours of these quantities analytically
as well as using numerical methods. In all cases, the $\text{E}_{W}$ decrease monotonically with temperature.
We observe discontinuous jumps in $\text{E}_{W}$ while the width of (as well as the separation between) the
subregions reach critical values in all the cases considered. However, the corresponding holographic mutual
information $\texttt{I}_{M}$ continuously decreases to zero for the aforementioned configurations. We also
notice that the conjectured inequality $\text{E}_{W} \geq \texttt{I}_{M}/2$ still holds for non-conformal field
theories as well. We analytically determine the critical separation between these subregions that triggers a
phase transition in the holographic mutual information.
\end{abstract}

\section{Introduction}\label{introduction}
The Gauge/Gravity duality \cite{Maldacena:1997re,Witten:1998qj,Aharony:1999ti} has enriched our understanding of
concepts related to quantum information theory. Perhaps the most prominent of all these examples is the holographic
computations of entanglement entropy of the dual boundary theory of interest which is particularly useful in determining
the entanglement entropy of pure states \cite{Ryu:2006bv,Ryu:2006ef,Hubeny:2007xt,Casini:2011kv,Faulkner:2013ana}.
This holographic entanglement entropy (HEE) proposal presents an interpretation of the entanglement entropy of the
boundary field theory in terms of a geometric quantity, namely, the minimal surface which is extended into the bulk.
However, this prescription can easily be extended to the case of more than one boundary intervals \cite{Ben-Ami:2014gsa}.
In this regard, a related quantity known as the holographic mutual information (HMI), $\texttt{I}_{M}\qty(A,B)$, between
two disjoint boundary intervals $A$ and $B$ can be determined \cite{Ben-Ami:2014gsa,Wolf:2007tdq,Headrick:2010zt, 
Fischler:2012uv}. Interestingly, $\texttt{I}_{M} \qty(A,B)\geq 0$ and it is UV finite. Moreover, it undergoes a phase transition
at some critical separation between the two intervals in which case the two subsystems become disentangled 
\cite{Ben-Ami:2014gsa,Headrick:2010zt}.

On the other hand, in order to determine the entanglement entropy of mixed states, an interesting measure known as the 
holographic entanglement wedge cross section (EWCS), $\text{E}_{W}$, is of much discussion in recent times 
\cite{Takayanagi:2017knl,Nguyen:2017yqw}. It is defined to be proportional to the minimal area of the entanglement wedge 
constructed out of the two intervals $A$ and $B$\footnote{The entanglement wedge is a bulk region whose boundary is given
by $\partial W_{AB}=A\cup B \cup \Gamma_{AB}^{m}$. Here $\Gamma_{AB}^{m}$ is the minimal surface associated with
the union $AB$.}. This quantity is conjectured to be holographic dual to entanglement of purification (EoP) \cite{Terhal:2002bm} 
which measures the correlation between $A$ and $B$ and, for pure states, reduces to entanglement entropy\footnote{It must
be mentioned that, the $\text{E}_{W}$ has also been proposed to be gravity dual of holographic entanglement negativity
\cite{Kudler-Flam:2018qjo} and reflected entropy \cite{Dutta:2019gen,Jeong:2019xdr}.}. Although this proposal still has not
been fully understood, there are certain indirect tests that this conjecture has passed, such as the following inequality 
\cite{Terhal:2002bm},
\begin{equation}\label{rel::EwMI}
\text{E}_{W} \geq \frac{\texttt{I}_{M}\qty(A,B)}{2},
\end{equation} 
which has been proven explicitly using the holographic duality \cite{Takayanagi:2017knl,Nguyen:2017yqw}. Very recently,
based on the formalism developed in \cite{Takayanagi:2017knl,Nguyen:2017yqw}, various properties of EWCS along with
HEE and HMI have been studied in different holographic set-ups \cite{BabaeiVelni:2019pkw,Jokela:2019ebz, 
Chakrabortty:2020ptb,Huang:2019zph,Boruch:2020wbe,Fu:2020oep}\footnote{Another promising direction, which is
extensively based on the entanglement wedge construction, is the study of time evolution of mixed state correlation
measures after global as well as local quantum quenches; see, for example, \cite{Yang:2018gfq,Kudler-Flam:2020url,
BabaeiVelni:2020wfl}. However, we shall not touch upon this aspect in this article.}. All these examples mostly deal with
conformal boundary theories in which cases the inequality (\ref{rel::EwMI}) holds explicitly. Moreover, both $\text{E}_{W}$
and HMI decrease monotonically and $\text{E}_{W}$ is shown to undergo a discontinuous phase transition at high
temperature.\footnote{Note that, in \cite{Jokela:2019ebz} non-monotonous behaviours of HMI and EWCS have been reported, 
however in a different holographic set-up.} 

The purpose of the present paper is to extend the study of certain aspects of EWCS, HEE and HMI to non-conformal field
theories which are dual to non-conformal D$p$-branes \cite{Itzhaki:1998dd,Boonstra:1998mp,Kanitscheider:2008kd,
Pang:2013lpa,Pang:2015lka}. This will eventually extend the formalism to the study of non-conformal field theories as well,
which is presently absent in the literature. This, in turn, could enable us to explore properties of scale non-invariant field
theories abundant in Nature.

The holographic correspondence between non-conformal D$p$-branes and their dual field theories were
studied in details in \cite{Itzhaki:1998dd} where the (super)gravity description is controlled by an effective dimensionless
coupling constant $g_{\text{eff}}$. However, in a more general set up, the 10-dimensional theory of \cite{Itzhaki:1998dd} can
be dimensionally reduced to obtain an effective $(p+2)$-dimensional theory described by Einstein-dilaton gravity 
\cite{Boonstra:1998mp,Kanitscheider:2008kd,Pang:2013lpa,Pang:2015lka}. Interestingly, the space-time metric obtained in
the later case is found to be conformal to that of AdS$_{p+2}$\footnote{Starting from \cite{Itzhaki:1998dd}, non-conformal
brane plane wave backgrounds have also been constructed in \cite{Narayan:2012hk,Singh:2012un} which are shown to
allow non-trivial hyperscaling violating backgrounds upon appropriate dimensional reduction. Also, aspects of holographic 
entanglement entropy have been studied within these theories; see, e.g., \cite{Narayan:2013qga,Mishra:2016yor}.}.

In this paper, we study the EWCS of two disjoint boundary intervals in the shape of parallel strips each of width $\ell$ and
length $L$. The low as well as high temperature behaviours of the EWCS are also studied. We restrict ourselves to the
cases $p = 1,2,4$. However, results corresponding to the conformal $p=3$ case are also provided for sake of completeness
of the analyses. We also discuss the behaviours of the HEE corresponding to the thermal boundary field theory along
the line of analysis of  \cite{Fischler:2012uv,vanNiekerk:2011yi,Fischler:2012ca,Kundu:2016dyk}. In our analysis we
observe that, in all cases the EWCS scales with the area of the entangling region contrary to the volume scaling of the
HEE. In addition, it monotonically decreases to zero value for large temperatures beyond certain critical value of the
separation $h$ between the two disjoint intervals at which they become disentangled. However, in all cases, the
corresponding HMI are monotonically decreasing functions as the critical separation between the strips is approached.
Moreover, we also check the validity of the above inequality (\ref{rel::EwMI}) explicitly in these set-ups.

The paper is organized as follows. In section \ref{set::up} we give a brief account of the non-conformal gravity backgrounds.
In section \ref{ewcs} we provide the computation of the EWCS and HEE corresponding to the strip shaped regions. The low
and high temperature behaviours of these quantities are studied subsequently. The values of the critical distance of
separation between the entangling regions are also determined by studying the corresponding HMI. We finally conclude in
section \ref{conclusion}.

\section{The holographic set up}\label{set::up}
In this paper, we study quantum information quantities for certain non-conformal field theories. In order to do so, we consider 
nonconformal D$p$-brane backgrounds dual to these field theories \cite{Itzhaki:1998dd}. Here we first mention the
corresponding constructions provided in \cite{Itzhaki:1998dd} and then very briefly review a more general dimensionally
reduced model following \cite{Boonstra:1998mp,Kanitscheider:2008kd} which we subsequently consider in all our computations.

The background generated from $N$ coincident extremal D$p$-branes in the string frame can be written as
\cite{Itzhaki:1998dd}
\begin{align}\label{NDp::gen}
\begin{split}
\dd s^{2}&=\alpha' \Bigg[ \frac{U^{\qty(7-p)/2}}{g_{YM}\sqrt{\dd_{p}N}}
\qty(-\dd t^{2}+\sum_{i=1}^{p}\dd x_{i}^{2})+\frac{g_{YM}\sqrt{\dd_{p}N}}
{U^{\qty(7-p)/2}}\dd U^{2}+g_{YM}\sqrt{\dd_{p}N}U^{\qty(p-3)/2}\dd
\Omega_{8-p}^{2}\Bigg],  \\
e^{\phi}&=\qty(2\pi)^{2-p}g_{YM}^{2}\qty(\frac{g_{YM}^{2}N\dd_{p}}
{U^{7-p}})^{\frac{3-p}{4}},
\end{split}
\end{align}
where 
\begin{equation}
\dd_{p}=2^{7-p}\pi^{\frac{9-3p}{2}}\Gamma\qty(\frac{7-p}{2}), \qquad
U=\frac{r}{\alpha'},
\end{equation}
and the limits $\displaystyle g_{YM}^{2}\sim g_{s}\alpha'^{\frac{p-3}{2}}=\text{fixed}$, $U=\text{fixed}$ and $\alpha' \to 0$
have been taken. On top of that, the validity of the supergravity description is controlled by the dimensionless coupling
constant $g_{eff}=g_{YM}^{2}NU^{p-3}$. In the non-extremal limit the above solution (\ref{NDp::gen}) can be expressed
as
\begin{align}\label{NEext::Dp}
\begin{split}
\dd s^{2}&=\alpha' \Bigg[ \frac{U^{\qty(7-p)/2}}{g_{YM}\sqrt{\dd_{p}N}}
\qty(-f\qty(U)\dd t^{2}+\sum_{i=1}^{p}\dd x_{i}^{2})+\frac{g\qty(U)
g_{YM}\sqrt{\dd_{p}N}}{U^{\qty(7-p)/2}}\dd U^{2}+g_{YM}\sqrt{\dd_{p}N}
U^{\qty(p-3)/2}\dd\Omega_{8-p}^{2}\Bigg],\\
e^{\phi}&=\qty(2\pi)^{2-p}g_{YM}^{2}\qty(\frac{g_{YM}^{2}N\dd_{p}}
{U^{7-p}})^{\frac{3-p}{4}},
\end{split}
\end{align}
where 
\begin{align}\label{MetCof::Dp}
\begin{split}
f\qty(U)=1-\frac{U_{0}^{7-p}}{U^{7-p}}, \qquad g\qty(U)=\frac{1}{f\qty(U)}
\approx1+\frac{U_{0}^{7-p}}{U^{7-p}}.
\end{split}
\end{align}

However, in a more general set up, we can perform an $S^{8-p}$ Kaluza-Klein compactification of the above
string-frame metric (\ref{NEext::Dp}) to $(p+2)$-dimensions. The effective theory (in the Einstein frame) is then
characterized by the $(p+2)$-dimensional Einstein-Dilaton action \cite{Boonstra:1998mp,Kanitscheider:2008kd,Pang:2013lpa,
Pang:2015lka}
\begin{align}\label{Recd::Dp}
\begin{split}
S=\frac{N^{2}}{16\pi G_{\mathsf{N}}^{p+2}}\qty[\int \dd^{p+2}x\sqrt{-g}\qty(\mathcal{R}
-\frac{1}{2}\partial_{\mu}\Phi\partial^{\mu}\Phi + V\qty(\Phi))-2\int \dd^{p+1}x
\sqrt{-\gamma}\mathcal{K}],
\end{split}
\end{align}
where
\begin{align}\label{Qts::Act}
V\qty(\Phi)&=\frac{1}{2}\qty(9-p)\qty(7-p)N^{-2\lambda/p}e^{a\Phi},&
\Phi&=\frac{2\sqrt{2\qty(9-p)}}{\sqrt{p}\qty(7-p)}\phi, \\
a&=-\frac{\sqrt{2}\qty(p-3)}{\sqrt{p\qty(9-p)}},& \lambda&=\frac{2
\qty(p-3)}{\qty(7-p)},
\end{align}
and $\mathcal{K}$, $\gamma_{ab}$ are the extrinsic curvature and the induced boundary metric, respectively. Also in
(\ref{Recd::Dp}) $G_{\mathsf{N}}^{p+2}$ is the $\qty(p+2)$-dimensional Newton's constant. The above theory (\ref{Recd::Dp}) 
allows for black brane solutions given by \cite{Boonstra:1998mp}-\cite{Pang:2015lka}
\begin{subequations}
\begin{equation}
\dd s^{2}=\qty(Ne^{\phi})^{\frac{2\lambda}{p}}\qty[\frac{u^{2}}{\Delta^{2}}
\qty(-f(u)\dd t^{2}+\sum_{i=1}^{p}\dd x_{i}^{2})+\frac{\Delta^{2}\dd u^{2}}
{u^{2}f(u)}],  \label{met::BB}
\end{equation}
\begin{equation}
f(u)=1-\qty(\frac{u_{0}}{u})^{\frac{2(7-p)}{(5-p)}},  \label{metcof::BB}
\end{equation}
\begin{equation}
e^{\phi}=\frac{1}{N}\qty(g_{YM}^{2}N)^\frac{7-p}{2(5-p)}\qty(\frac{u}{\Delta})^{\frac{(p-7)
(p-3)}{2(p-5)}},  \label{diln::BB} 
\end{equation}
\begin{equation}
u_{0}^{2}=\frac{U_{0}^{5-p}}{\qty(g_{YM}^{2}N)},\qquad \Delta=\frac{2}{5-p}.
\label{hozn::BB}
\end{equation}
\end{subequations}

Now the Hawking temperature of the D$p$-brane can be computed by the usual method of analytical continuation of the
metric (\ref{met::BB}) to the Euclidean sector, $t \to i\tau$ \cite{Wald:1984}. The resulting expression for the
temperature can be calculated as
\begin{equation}\label{temp::Hkg}
T=\frac{1}{4\pi \Delta^{2}}\frac{2\qty(p-7)}{\qty(p-5)z_{0}},
\end{equation}
where the following change in coordinates $\displaystyle u\to 1/z$ and $u_{0}\to 1/z_{0}$ has been taken into account
\cite{Pang:2013lpa,Pang:2015lka}.

On the other hand, the thermal entropy of the D$p$-branes can be computed as
\begin{align}\label{Ent::thml}
\begin{split}
S_{\text{th}}&=\frac{1}{4G_{\mathsf{N}}^{p+2}}\int_{-\ell/2}^{\ell/2}\dd x \int_{-L/2}^{L/2}
\prod_{i=1}^{p-1}\dd x_{i} \qty(g_{xx}\prod_{i=1}^{p-1}g_{x_{i}x_{i}})^{\frac{1}{2}}
\\
&=\frac{L^{p-1}\ell}{4G_{\mathsf{N}}^{p+2}}\qty(g_{YM}^{2}N)^{\frac{p-3}{5-p}}
\qty(\frac{1}{z_{0}\Delta})^{\frac{p-9}{p-5}}.
\end{split}
\end{align}


\section{Entanglement Wedge Cross Section (EWCS)}\label{ewcs}
\begin{figure}[t]
\centering
\begin{tikzpicture}
\draw[very thick, blue] (0,0) arc (0:-180:3cm);
\draw[very thick, red] (-2.2,0) arc (0:-180:0.8cm);

\filldraw[very thick,black] (0,0) -- (-2.2,0);
\filldraw[very thick,black] (-3.8,0) -- (-6,0);
\filldraw[very thick,green] (-3,-0.8) -- (-3,-3);
\filldraw[very thick, dashed] (-7,-3.65) -- (1,-3.65);
\filldraw[thick, dashed] (-7,0) -- (1,0);
\draw[|-|]  (-2.2,0.3) -- node[above]{$h$} (-3.8,0.3);
\draw[<->]  (0,0.3) -- node[above]{$\ell$} (-2.2,0.3);
\draw[<->]  (-3.8,0.3) -- node[above]{$\ell$} (-6,0.3);

\node at (-1.1,-0.3) {$ A $};
\node at (-4.5,-0.3) {$ B $};
\node at (1.25,-3.65) {$ z_{0} $};
\node at (1.25,0) {$ x $};
\node at (-2.5,-2.2) {$ \Gamma_{AB}^{m} $};

\end{tikzpicture}
\caption{A schematic 2D diagram for computing entanglement wedge
cross section (EWCS) $\Gamma_{AB}^{m}$. Here we consider two
parallel strips each of width $\ell$ separated by $h$. Here $z_{0}$
represents the horizon of the brane. }
\label{Fig:Ew}
\end{figure}
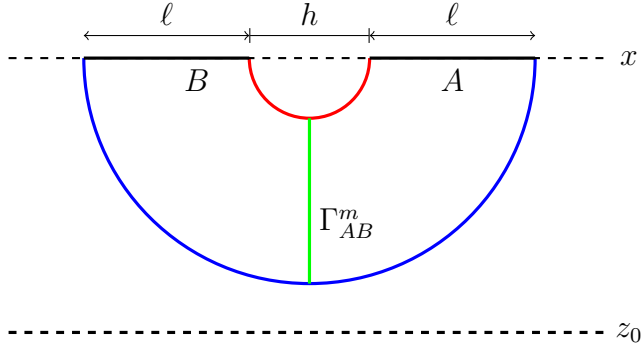

Let us consider two long parallel strips each of width $\ell$ and separated by a distance $h$. They indeed represent two
subregions $A$ and $B$ and is shown as dark black lines in fig.\ref{Fig:Ew}. We choose the following specific symmetric
(around $x=0$) configurations for the subsystems \cite{BabaeiVelni:2019pkw,Jokela:2019ebz,Chakrabortty:2020ptb}:
\begin{subequations}
\setlength{\jot}{8pt}
\begin{eqnarray}
A&=&\qty{\frac{h}{2}<x<\ell+\frac{h}{2};\qquad -\frac{L}{2}
<x_{2},x_{3},\cdots,x_{p-1}<\frac{L}{2}},    \label{conf:R1}    \\
B&=&\qty{-\ell-\frac{h}{2}<x<-\frac{h}{2};\quad -\frac{L}{2}
<x_{2},x_{3}\cdots,x_{p-1}<\frac{L}{2}}    \label{conf:R1}.
\end{eqnarray}
\end{subequations}

Now the minimal surface, $\Gamma^{m}_{AB}$, that separates the two subsystems $A$ and $B$ is given by the vertical
constant $x$ surface at $x=0$ which is a space-like slice. The induced metric on this slice can be written as
\begin{align}\label{ind:met:gma}
\begin{split}
\dd s_{\Gamma^{m}_{AB}}^{2}&=\qty(Ne^{\phi})^{\frac{2\lambda}{p}}\qty[\frac{1}
{z^{2}\Delta^{2}}\sum_{i=1}^{p-1}\dd x_{i}^{2}+\frac{\Delta^{2}\dd z^{2}}
{z^{2}f(z)}], \\
f(z)&=1-\qty(\frac{z}{z_{0}})^{\frac{2(7-p)}{(5-p)}}.
\end{split}
\end{align}

Next, we use the general expression for the entanglement wedge cross section (EWCS) corresponding to the two
subregions $A$ and $B$ given by 
\cite{Takayanagi:2017knl,Nguyen:2017yqw,BabaeiVelni:2019pkw,Jokela:2019ebz,Chakrabortty:2020ptb}
\begin{equation}\label{ewcs::gen}
\text{E}_{W}=\frac{\text{Area}\qty(\Gamma^{m}_{AB})}
{4G_{\mathsf{N}}^{p+2}}.
\end{equation}

Using (\ref{ind:met:gma}) and (\ref{ewcs::gen}) we finally obtain\footnote{Notice that the integration in (\ref{ewcs::expn})
can be evaluated exactly; however, we find it convenient to express it in the series form \cite{Fischler:2012uv,Fischler:2012ca}.}
\begin{align}\label{ewcs::expn}
\begin{split}
\text{E}_{W}&=\frac{L^{p-1}\qty(g_{YM}^{2}N)^{\frac{p-3}{5-p}}}{4G_{\mathsf{N}}^{p+2}
\Delta^{\frac{p-1}{5-p}}}\int_{z_{t}(h)}^{z_{t}(2\ell+h)}\dd z \frac{z^{\frac{9-p}
{p-5}}}{\sqrt{f(z)}}\\
&=\frac{L^{p-1}\qty(g_{YM}^{2}N)^{\frac{p-3}{5-p}}}{4G_{\mathsf{N}}^{p+2}\Delta^{\frac{p-1}{5-p}}}
\sum_{n=0}^{\infty}\frac{\Gamma\qty(n+\frac{1}{2})\qty(p-5)}{\sqrt{\pi}\qty(4+2n\qty(p-7))
\Gamma\qty(n+1)}\qty(\frac{z_{t}\qty(2\ell+h)^{\delta}}{z_{0}^{n\alpha}}-
\frac{z_{t}\qty(h)^{\delta}}{z_{0}^{n\alpha}}),
\end{split}
\end{align}
where we have denoted
\begin{equation}
\alpha = \frac{2(7-p)}{(5-p)}, \qquad \qquad \delta=\frac{4+2n\qty(p-7)}{(p-5)}.
\end{equation}

We now use the Ryu-Tagayanagi prescription \cite{Ryu:2006bv,Ryu:2006ef} in order to find the holographic entanglement
entropy (HEE) corresponding to a strip shaped region in the boundary of the geometry (\ref{met::BB}) which can be
expressed as \cite{vanNiekerk:2011yi}
\begin{equation}\label{HEE:gexpn}
S_{\text{EE}}=\frac{\mathcal{A}_{\gamma}}{4G_{\mathsf{N}}^{p+2}},
\end{equation}
where $\mathcal{A}_\gamma$ is the area of the $p$-dimensional Ryu-Takayanagi surface $\gamma$.

In our analysis, we choose the strip parametrized as 
\begin{equation}\label{geom::HEE}
-\frac{\ell}{2} \leq x \leq \frac{\ell}{2}, \qquad \qquad -\frac{L}{2} \leq x_{2},\cdots, x_{p-1}
\leq \frac{L}{2}.
\end{equation}

Thus the HEE functional can be wriiten as
\begin{equation}\label{HEE:Dp}
S_{\text{EE}}=\frac{L^{p-1}}{2G_{\mathsf{N}}^{p+2}}\qty(g_{YM}^{2}N)^{\frac{p-3}{5-p}}
\Delta^{\frac{9-p}{p-5}}\int \dd z \;z^{\frac{9-p}{p-5}}\sqrt{x'^{2}+\frac{\Delta^{4}}
{f(z)}}.
\end{equation}

Interestingly, the above functional (\ref{HEE:Dp}) has no explicit dependence on $x(z)$, and hence we can find the
following first integral of motion by applying the conservation of energy of the system,
\begin{equation}\label{IM::covd}
x'(z)=\pm \frac{\Delta^{2}}{\sqrt{f(z)\qty(\qty(\frac{z}{z_{t}})^{2\frac{(9-p)}{(p-5)}}-1)}},
\end{equation}
where $z_{t}$ is the turning point of the minimal surface. Also note that, in deriving (\ref{IM::covd}) we have used the
boundary condition $\displaystyle \lim_{x\to\infty}z=z_{t}$. 

In the next step, we integrate (\ref{IM::covd}) to obtain a relation between strip width $\ell$ and the turning point $z_{t}$
as
\begin{equation}\label{Int::wdth}
\ell = 2 \Delta^{2} z_{t} \int_{0}^{1} \dd v \frac{v^{\frac{9-p}{5-p}}}{\sqrt{f(v)}
\sqrt{1-v^{2\frac{(9-p)}{(5-p)}}}},
\end{equation}
where we have defined $v=\frac{z}{z_{t}}$ and considered the fact that $1\leq p \leq 4$.

Substituting (\ref{IM::covd}) into (\ref{HEE:Dp}) we finally obtain
\begin{equation}\label{HEE:02}
S_{\text{EE}}=\frac{L^{p-1}}{2G_{\mathsf{N}}^{p+2}}\qty(g_{YM}^{2}N)^{\frac{p-3}{5-p}}
\frac{\Delta^{\frac{p-1}{p-5}}}{z_{t}^{\frac{4}{5-p}}}\int_{0}^{1} \frac{\dd v}
{v^{\frac{9-p}{5-p}}}\frac{1}{\sqrt{f(v)\qty(1-v^{2\frac{(9-p)}{(5-p)}})}}.
\end{equation}

The integrations appearing in the above equations (\ref{Int::wdth}) and (\ref{HEE:02}) can be evaluated easily; and
hence the width and the HEE can respectively be expressed as \cite{Fischler:2012uv,Fischler:2012ca}
\begin{equation}\label{wdth::finl}
\ell =\Delta^{2}z_{t}\sum_{n=0}^{\infty}\frac{\Gamma\qty(n+\frac{1}{2})
\Gamma\qty(\frac{(n+1)(p-7)}{p-9})}{\Gamma\qty(n+1)\Gamma
\qty(\frac{3}{2}+n+\frac{2(n+1)}{(p-9)})}\frac{(p-5)}{(p-9)}
\qty(\frac{z_{t}}{z_{0}})^{2n\frac{(7-p)}{(5-p)}},
\end{equation}

\begin{align}\label{HEE::finl}
\begin{split}
S_{\text{EE}}=S_{\text{sin}}+&\frac{L^{p-1}}{2G_{\mathsf{N}}^{p+2}}\qty(g_{YM}^{2}N)^{\frac{p-3}{5-p}}
\frac{\Delta^{\frac{p-1}{p-5}}}{z_{t}^{\frac{4}{5-p}}}\Bigg[\frac{\sqrt{\pi}\qty(p-5)
\Gamma\qty(\frac{2}{p-9})}{2\qty(p-9)\Gamma\qty(\frac{p-5}{2(p-9)})}  \\
&+\sum_{n=1}^{\infty}\frac{\Gamma\qty(n+\frac{1}{2})\Gamma\qty(n+\frac{2(n+1)}
{p-9})}{\Gamma\qty(n+1)\Gamma\qty(\frac{p-5+2n(p-7)}{2(p-9)})}\frac{(p-5)}{2(p-9)}
\qty(\frac{z_{t}}{z_{0}})^{2n\frac{(7-p)}{(5-p)}}\Bigg],
\end{split}
\end{align}
where the singular part of the HEE can be written as
\begin{equation}\label{HEE::singr}
S_{\text{sin}}=\frac{L^{p-1}}{2G_{\mathsf{N}}^{p+2}}\qty(g_{YM}^{2}N)^{\frac{p-3}{5-p}}
\frac{\Delta^{\frac{p-1}{p-5}}}{\epsilon^{\frac{4}{5-p}}}\qty(\frac{5-p}{4}).
\end{equation}
Notice that, in (\ref{HEE::singr}) $\epsilon$ is the ultraviolet (UV) cut-off.

At this point of discussion, note that, we can get the expressions for $E_{W}$ as well as $S_{\text{EE}}$ corresponding to
the conformal (D$3$-brane) case by substituting $p=3$ in (\ref{ewcs::expn}) and (\ref{HEE:02}), respectively. In the next
sections we also compute the corresponding expressions for D$3$-branes for completeness of the analysis which indeed
match with those already exist in the literature; see, e.g., 
\cite{Ryu:2006ef,Ben-Ami:2014gsa,BabaeiVelni:2019pkw,Jokela:2019ebz,Chakrabortty:2020ptb,Fischler:2012ca,Kundu:2016dyk}. 
In addition, we shall also study the corresponding quantities by plotting their behaviours numerically. In our numerical
analysis, we shall set the pre factor $\frac{L^{p-1}}{4G_{\mathsf{N}}^{p+2}}\frac{\qty(g_{YM}^{2}N)^{\frac{(p-3)}{(5-p)}}}
{\Delta^{\frac{p-1}{5-p}}}=1$.

\begin{figure}[t!]
    \centering
    {{\includegraphics[width=8cm]{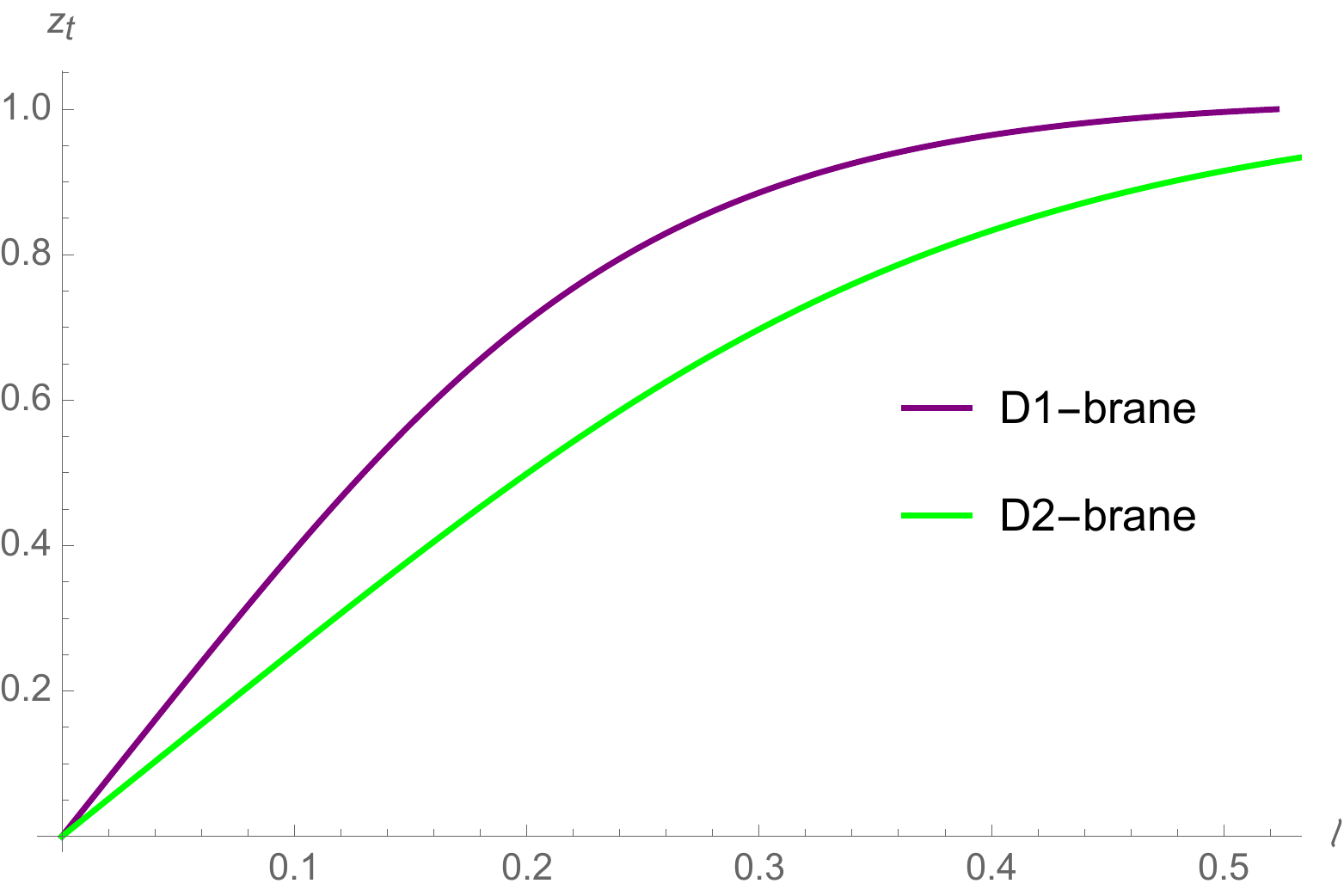} }}%
    \qquad
    {{\includegraphics[width=8cm]{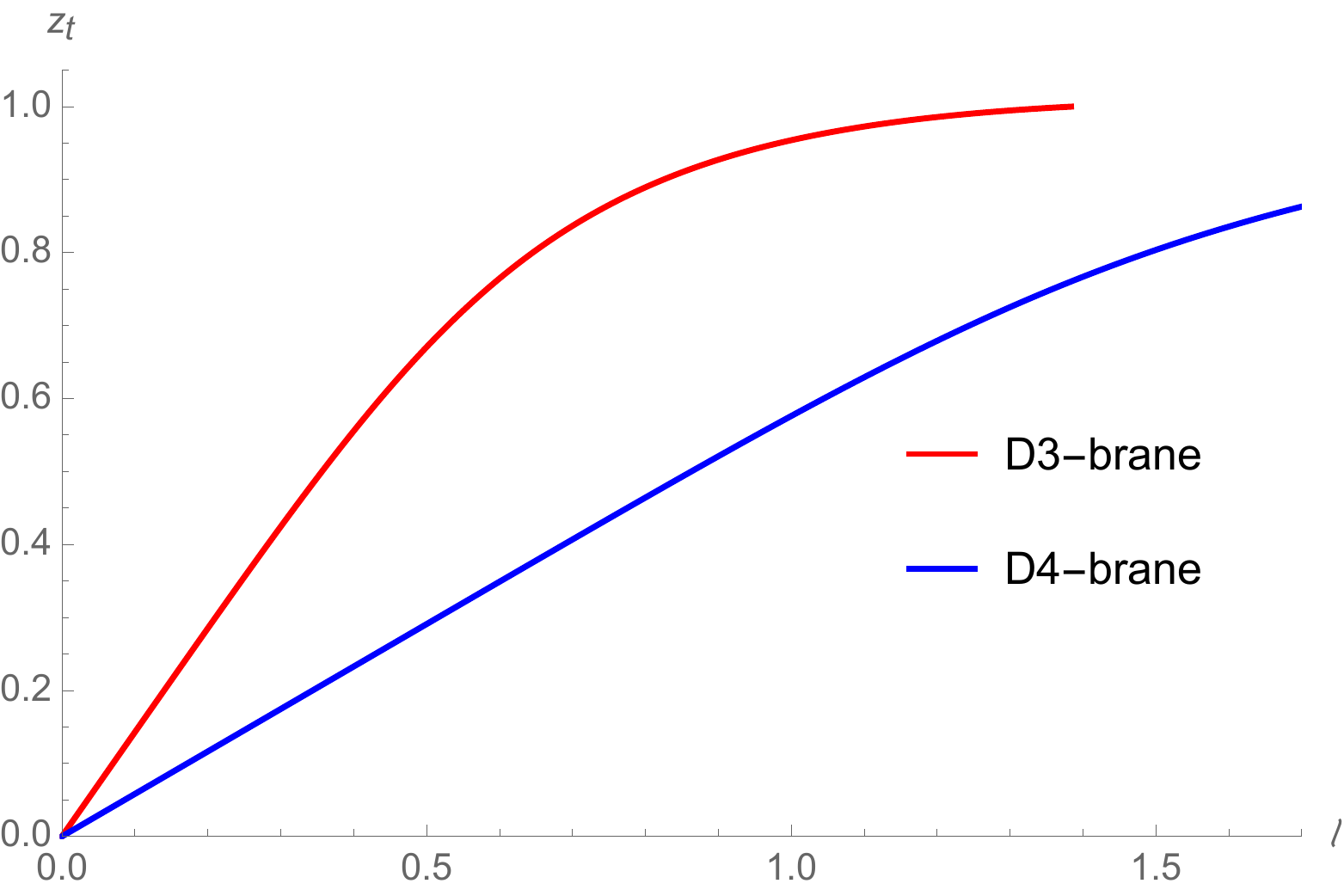} }}%
\caption{$z_{t}$ \emph{vs.} $\ell$ plot for different D$p$-brane backgrounds.
The turning point vanishes at $\ell=0$, and as $\ell$ increases, $z_{t}$
approaches 1. Here we set the horizon radius $z_{0}=1$.}
\label{Fig:1}
\end{figure}

\subsection{Low and high temperature behaviours of the HEE}\label{LT::HEE}
In this section, we study both the low and high temperature behaviours of the holographic entanglement entropy (HEE) 
(\ref{HEE::finl}) in the backgrounds (\ref{met::BB}). In order to achieve this analytically, we need to solve the turning point
$z_{t}$ in terms of the subregion length $\ell$. It is evident from (\ref{wdth::finl}) that this procedure can only be applied in
the low and high temperature limits.

The low temperature limit is geometrically realized when the turning point $z_{t}$ corresponding to the RT surface(s) lies
far away from the horizon $z_{0}$ of the black brane in the deep interior of the bulk: $z_{t} \ll z_{0}$. On the other hand, for
the considered range of values of $p$, $1\leq p \leq 4$, the exponent $\frac{(7-p)}{(5-p)}\leq 3$. Hence, it is sufficient to keep
terms upto order $\order{4\frac{(7-p)}{(5-p)}}$ in the perturbative expansion. Thus the resulting expression for the turning
point in (\ref{wdth::finl}) may be written as
\begin{align}\label{turn::pt}
\begin{split}
z_{t} \approx \frac{\ell}{\Delta^{2}\Upsilon}&\left[ 1- \frac{\sqrt{\pi}(p-5)
\Gamma\qty(\frac{2(p-7)}{p-9})}{2(p-9)\Gamma\qty(\frac{5}{2}+\frac{4}{p-9})}
\frac{1}{\Upsilon}\qty(\frac{\ell}{\Delta^{2}z_{0}\Upsilon})^{\frac{2(7-p)}{(5-p)}}
+\left\{ \qty(\frac{\sqrt{\pi}(p-5)\Gamma\qty(\frac{2(p-7)}{p-9})}{2(p-9)\Gamma
\qty(\frac{5}{2}+\frac{4}{p-9})}\frac{1}{\Upsilon})^{2} \right.\right. \\
&\left.\left. -\qty(\frac{3\sqrt{\pi}(p-5)}{8(p-9)}\frac{\Gamma\qty(\frac{3(p-7)}{(p-9)})}
{\Gamma\qty(\frac{7}{2}+\frac{6}{p-9})}\frac{1}{\Upsilon})\right\}\qty(\frac{\ell}
{\Delta^{2}z_{0}\Upsilon})^{\frac{4(7-p)}{(5-p)}}+\order{\frac{\ell}{\Delta^{2}
z_{0}\Upsilon}}^{\frac{6(7-p)}{(5-p)}}\right],
\end{split}
\end{align}
where we have defined
\begin{equation}
\Upsilon = \frac{\sqrt{\pi}(p-5)\;\Gamma\qty(\frac{p-7}{p-9})}
{(p-9)\;\Gamma\qty(\frac{3}{2}+\frac{2}{p-9})}.
\end{equation}

With this approximation and substituting (\ref{turn::pt}) into (\ref{HEE::finl}), the final expressions for the HEE at low
temperatures can be written as follows.\footnote{Here and what follows, the subscripts `\textit{LT}' and `\textit{HT}' denote
Low Temperature and High Temperature, respectively.}

\begin{flushleft}
\textbf{\underline{D1-brane}}
\end{flushleft}

\begin{equation}\label{HEED1::LT}
S_{\text{EE}_{LT}}^{(D1)}=\bar{\mathsf{C}_{1}}\qty[S_{0}+\mathcal{C}_{1}
\qty(\frac{1}{\ell}) \qty(1+\mathcal{C}_{2}\qty(\frac{\pi T \ell}{3})^{3}
+\mathcal{C}_{3}\qty(\frac{\pi T \ell}{3})^{6}+\order{\pi T \ell/3}^{7})],
\end{equation}
where we have denoted
\begin{align}
\bar{\mathsf{C}_{1}}&=\frac{\qty(g_{YM}^{2}N)^{-\frac{1}{2}}}{4G_{\mathsf{N}}^{3}},&
S_{0}&=\frac{2}{\epsilon},
\end{align}
and the other \emph{constants} $\mathcal{C}_{1}$, $\mathcal{C}_{2}$ and $\mathcal{C}_{3}$ are constants. We have
explicitly written down the forms of these constants and those that appear below ($\mathcal{D}_{i}$, $\mathcal{B}_{i}$
and $\mathcal{E}_{i}$, $i=1,2,3$) in appendix \ref{coefs}.

\begin{flushleft}
\textbf{\underline{D2-brane}}
\end{flushleft}

\begin{equation}\label{HEED2::LT}
S_{\text{EE}_{LT}}^{(D2)}=\bar{\mathsf{C}_{2}}\qty[S_{0}+\mathcal{D}_{1}
\qty(\frac{1}{\ell})^{4/3} \qty(1+\mathcal{D}_{2}\qty(\frac{8\pi T\ell}{15})^{10/3}
+\mathcal{D}_{3}\qty(\frac{8\pi T \ell}{15})^{20/3}+\order{8\pi T\ell/15}^{19/3})],
\end{equation}
where
\begin{align}
\bar{\mathsf{C}_{2}}&=\frac{L\qty(g_{YM}^{2}N)^{-\frac{1}{3}}}{4G_{\mathsf{N}}^{4}
\Delta^{\frac{1}{3}}},&
S_{0}&=\qty(\frac{3}{2})\qty(\frac{1}{\epsilon})^{4/3},
\end{align}
and $\mathcal{D}_{1}$, $\mathcal{D}_{2}$ and $\mathcal{D}_{3}$ are constants.

\begin{flushleft}
\textbf{\underline{D3-brane}}\cite{Ryu:2006ef,
Ben-Ami:2014gsa,BabaeiVelni:2019pkw,Jokela:2019ebz,Chakrabortty:2020ptb,Fischler:2012ca,Kundu:2016dyk}
\end{flushleft}

\begin{equation}\label{HEED3::LT}
S_{\text{EE}_{LT}}^{(D3)}=\bar{\mathsf{C}_{3}}\qty[S_{0}+\mathcal{B}_{1}
\qty(\frac{1}{\ell})^{2} \Bigg(1+\mathcal{B}_{2}\qty(\pi T\ell)^{4}
+\mathcal{B}_{3}\qty(\pi T \ell)^{8}+\order{\pi T\ell}^{10}\Bigg)],
\end{equation}
where
\begin{align}
\bar{\mathsf{C}_{3}}&=\frac{L^{2}}{4G_{\mathsf{N}}^{5}},    &
S_{0}&=\frac{1}{\epsilon^{2}},
\end{align}
and $\mathcal{B}_{1}$, $\mathcal{B}_{2}$ and $\mathcal{B}_{3}$ are constants.

\begin{flushleft}
\textbf{\underline{D4-brane}}
\end{flushleft}

\begin{equation}\label{HEED4::LT}
S_{\text{EE}_{LT}}^{(D4)}=\bar{\mathsf{C}_{4}}\qty[S_{0}+\mathcal{E}_{1}
\qty(\frac{1}{\ell})^{4} \qty(1+\mathcal{E}_{2}\qty(\frac{8\pi T\ell}{3})^{6}
+\mathcal{E}_{3}\qty(\frac{8\pi T \ell}{3})^{12}+\order{8\pi T\ell/3}^{18})],
\end{equation}
with
\begin{align}
\bar{\mathsf{C}_{4}}&=\frac{L^{3}\qty(g_{YM}^{2}N)}{4G_{\mathsf{N}}^{6}
\Delta^{3}},&
S_{0}&=\qty(\frac{1}{2})\qty(\frac{1}{\epsilon})^{4},
\end{align}
and $\mathcal{E}_{1}$, $\mathcal{E}_{2}$ and $\mathcal{E}_{3}$ are constants.

Notice that, when the temperature is zero ($T=0$), the leading order finite terms appearing in (\ref{HEED1::LT}),
(\ref{HEED2::LT}), (\ref{HEED3::LT}) and (\ref{HEED4::LT}) agree with those prescribed in \cite{vanNiekerk:2011yi}.
\hfill \break

On the other hand, at high temperatures the turning point $z_{t}$ approaches the horizon of the black branes, $z_{t}
\to z_{0}$, and indeed wraps a part of the horizon. Thus the leading contribution comes from the near horizon part of the
surface. On top of that, the entire bulk geometry contributes in the form of subleading terms \cite{Fischler:2012uv,
Fischler:2012ca}. The following combination is found to be converging as the limit $z_{t} \to z_{0}$ is taken and hence we
proceed our analysis with this combination \cite{Fischler:2012uv,Chakrabortty:2020ptb,Fischler:2012ca}.
\begin{align}\label{HEE:convg}
\begin{split}
S_{\text{EE}}-\frac{L^{p-1}\qty(g_{YM}^{2}N)^{\frac{p-3}{5-p}}}{4G_{\mathsf{N}}^{p+2}}
\qty(\frac{1}{\Delta z_{t}})^{\frac{9-p}{5-p}} \ell = \frac{L^{p-1}\qty(g_{YM}^{2}
N)^{\frac{p-3}{5-p}}}{2G_{\mathsf{N}}^{p+2}\Delta^{\frac{p-1}{5-p}}}
\qty(\frac{1}{z_{t}})^{\frac{4}{5-p}} \int_{\epsilon/z_{t}}^{1}\dd v \frac{\sqrt{1-
v^{\frac{2(9-p)}{5-p}}}}{\sqrt{f\qty(v)}\;v^{\frac{9-p}{5-p}}}.
\end{split}
\end{align}

Using (\ref{HEE:convg}) we can now recast the finite part of the HEE as
\begin{align}\label{HEE:new}
\begin{split}
S_{\text{EE}}=\frac{L^{p-1}\qty(g_{YM}^{2}N)^{\frac{p-3}{5-p}}}{4G_{\mathsf{N}}^{p+2}}
\qty(\frac{1}{\Delta z_{t}})^{\frac{9-p}{5-p}} \ell & + \frac{L^{p-1}\qty(g_{YM}^{2}
N)^{\frac{p-3}{5-p}}}{2G_{\mathsf{N}}^{p+2}\Delta^{\frac{p-1}{5-p}}}
\qty(\frac{1}{z_{t}})^{\frac{4}{5-p}} \Bigg[\frac{(p-5)}{4} \, _2F_1\left(\frac{1}{2},
\frac{2}{p-9};\frac{p-7}{p-9};1\right)  \\
& + \int_{0}^{1}\dd v \left( \frac{\sqrt{1-v^{\frac{2(9-p)}{5-p}}}}{\sqrt{f\qty(v)}\;
v^{\frac{9-p}{5-p}}} - \frac{1}{v^{\frac{9-p}{5-p}}\sqrt{1-v^{\frac{2(9-p)}{5-p}}}}
\right) \Bigg],
\end{split}
\end{align}
where $_2F_1\qty(a,b;c;z)$ is the usual hypergeometric function.

Finally, considering the limit $z_{t} \to z_{0}$, we can read off the expression for the HEE at high temperatures from 
(\ref{HEE:new}) as \cite{Fischler:2012uv,Fischler:2012ca}
\begin{align}\label{HEE::HT}
\begin{split}
S_{\text{EE}_{HT}}&=S_{\text{sin}}+\frac{V \qty(g_{YM}^{2}N)^{\frac{p-3}{5-p}}}{4G_{\mathsf{N}}^{p+2}
\Delta^{\frac{9-p}{5-p}}}\qty(\frac{2\pi\Delta^{2}(p-5)}{(p-7)}T)^{\frac{9-p}{5-p}}\Bigg\{ 1+
2\qty(\frac{(p-7)}{2\pi \ell T (p-5)})\Theta\Bigg\},
\end{split}
\end{align}
where $S_{\text{sin}}$ is the singular part of the HEE given by (\ref{HEE::singr}) and
\begin{align}\label{Theta}
\begin{split}
\Theta&=\frac{\sqrt{\pi}(p-5)}{(p-9)}\qty[\frac{\Gamma\qty(\frac{2}{p-9})}{2\Gamma
\qty(\frac{p-5}{2(p-9)})}-\frac{\Gamma\qty(\frac{p-7}{p-9})}{\Gamma\qty(\frac{3}{2}
+\frac{2}{(p-9)})}]  \\
&\quad+\sum_{n=1}^{\infty}\frac{(p-5)\Gamma\qty(n+\frac{1}{2})}{(p-9)\Gamma\qty(n+1)}
\qty[\frac{\Gamma\qty(n+\frac{2(n+1)}{p-9})}{2\Gamma\qty(\frac{p-5+2n(p-7)}{2(p-9)})}
-\frac{\Gamma\qty(\frac{(n+1)(p-7)}{p-9})}{\Gamma\qty(\frac{3}{2}+n+\frac{2(n+1)}{(p-9)})}].
\end{split}
\end{align}

Note that, the finite term of the HEE in (\ref{HEE::HT}) scales as the volume of the entangling region $V=\ell L^{p-1}$.
Also, this term is proportional to the thermal entropy (\ref{Ent::thml}). This is expected since at high temperature regime
the contributions to the HEE of the thermal boundary field theory comes from the thermal fluctuations. Furthermore,
when $p=3$ the HEE at high temperature, (\ref{HEE::HT}), corresponds to the conformal case as discussed in 
\cite{Ryu:2006ef,Ben-Ami:2014gsa,BabaeiVelni:2019pkw, Jokela:2019ebz,Chakrabortty:2020ptb,Fischler:2012ca,
Kundu:2016dyk}.

\subsection{Low and high temperature behaviours of the EWCS}\label{LT::EWCS}
In order to study the low and high temperature behaviours of the EWCS, we first notice that we can achieve this by taking
into account the following three intrinsic scales associated with the theory: the separation distance $h$ between the two
entangling regions, the width $\ell$ of each of the regions and the temperature $T$ of the boundary theory 
\cite{Fischler:2012uv,BabaeiVelni:2019pkw,Jokela:2019ebz,Chakrabortty:2020ptb,Fischler:2012ca}. Subsequently, in this
case the low temperature limit corresponds to $hT \ll \ell T\ll 1$. Note that, this corresponds to considering the temperature
smaller than both the length scales associated with $h$ and $\ell$. On the other hand, the high temperature limit may be
defined by considering the following inequality $hT \ll 1\ll \ell T$ in which case the temperature is large compared to the
scale associated with $\ell$ but small compared to that associated with $h$.\footnote{In our calculations we shall \emph{not}
take into account another limit $\ell T \ll h T$ or $1 \ll \ell T$, $1 \ll h T$ as this corresponds to the disentangling phase of the
two subregions \cite{Fischler:2012uv,BabaeiVelni:2019pkw,Chakrabortty:2020ptb,Fischler:2012ca,Kundu:2016dyk}.}
Interestingly, this high temperature limit further amounts to taking the following two approximations: $z_{t} \qty(h) \ll z_{0}$
and $z_{t} \qty(2 \ell + h) \to z_{0}$. The first approximation amounts to considering only the leading order term in the second
term within the braces in (\ref{ewcs::expn}), while in order the second approximation to be valid we must ensure the
convergence of the first sum in (\ref{ewcs::expn}). It is trivial to check that this is indeed the case. Thus it is safe to consider
the limit $z_{t}\qty(2\ell+h)\to z_{0}$.

Let us now discuss the different cases corresponding to $p=1,2,4$ separately which correspond to D1-, D2- and D4-branes, 
respectively. We have also provided the corresponding expressions for D$3$-branes (see, e.g.,
\cite{BabaeiVelni:2019pkw,Jokela:2019ebz,Chakrabortty:2020ptb}) for completeness.

\begin{flushleft}
\textbf{\underline{D1-brane}}
\end{flushleft}

In the low temperature limit $hT \ll \ell T\ll 1$ the EWCS (\ref{ewcs::expn}) for the D1-branes can be computed as
\begin{align}\label{EwD1::LT}
\begin{split}
\text{E}_{W_{LT}}^{(D1)}&=\frac{\qty(g_{YM}^{2}N)^{-\frac{1}{2}}}{4G_{\mathsf{N}}^{3}}\sum_{n=0}^{\infty}
\frac{\Gamma\qty(n+\frac{1}{2})}{\sqrt{\pi}\Gamma\qty(n+1)}\qty(\frac{z_{t}\qty(2\ell+h)^{3n-1}
-z_{t}(h)^{3n-1}}{3n-1})\cdot \frac{1}{z_{0}^{3n}}  \\
&=\frac{\qty(g_{YM}^{2}N)^{-\frac{1}{2}}}{4G_{\mathsf{N}}^{3}}\qty[\frac{\sqrt{\pi}\Gamma\qty(\frac{3}{4})}
{8\Gamma\qty(\frac{5}{4})}\qty(\frac{1}{h}-\frac{1}{2\ell +h})+\frac{64\;\Gamma\qty(\frac{5}{4})^{2}}
{\pi\Gamma\qty(\frac{3}{4})^{2}}\qty(1-\frac{\sqrt{\pi}\Gamma\qty(\frac{5}{4})}{\Gamma\qty(\frac{3}{4})})
\; \ell\qty(\ell +h)\qty(\frac{\pi T}{3})^{3} + \cdots],
\end{split}
\end{align}
where we have used (\ref{temp::Hkg}) and (\ref{turn::pt}).

On the other hand, considering the aforementioned high temperature limits the behaviour of the EWCS can be determined
as
\begin{align}\label{EwD1::HT}
\begin{split}
\text{E}_{W_{HT}}^{(D1)}\approx\frac{\qty(g_{YM}^{2}N)^{-\frac{1}{2}}}{4G_{\mathsf{N}}^{3}}
\;T\qty[\frac{\pi\tilde{\mathcal{C}}_{1}}{3}+\frac{\sqrt{\pi}\Gamma\qty(\frac{3}{4})}{8
\Gamma\qty(\frac{5}{4})}\qty(\frac{1}{hT})+\frac{16\pi^{\frac{5}{2}}}{27}\qty(\frac{
\Gamma\qty(\frac{5}{4})}{\Gamma\qty(\frac{3}{4})})^{3}\qty(hT)^{2}],
\end{split}
\end{align}
where we have defined $\displaystyle \tilde{\mathcal{C}}_{1}=\sum_{n=0}^{\infty}\frac{\Gamma\qty(n+\frac{1}{2})}{\sqrt{\pi}
\qty(3n-1)\Gamma\qty(n+1)}$. 

\begin{flushleft}
\textbf{\underline{D2-brane}}
\end{flushleft}

In a similar manner, the low temperature behaviour of EWCS for the D2-branes (\ref{ewcs::expn}) can be determined as
\begin{align}\label{EwD2::LT}
\begin{split}
\text{E}_{W_{LT}}^{(D2)}&=\frac{L\qty(g_{YM}^{2}N)^{-\frac{1}{3}}}{4G_{\mathsf{N}}^{4}
\Delta^{\frac{1}{3}}}\sum_{n=0}^{\infty}\frac{\Gamma\qty(n+\frac{1}{2})}
{\sqrt{\pi}\Gamma\qty(n+1)}\qty(\frac{z_{t}\qty(2\ell+h)^{\frac{10n-4}{3}}
-z_{t}(h)^{\frac{10n-4}{3}}}{10n-4})\cdot \frac{3}{z_{0}^{\frac{10}{3}n}}  \\
&=\frac{L\qty(g_{YM}^{2}N)^{-\frac{1}{3}}}{4G_{\mathsf{N}}^{4}\Delta^{\frac{1}{3}}}
\Bigg[\frac{3\qty(2\pi)^{\frac{2}{3}}}{\qty(21)^{\frac{4}{3}}}\qty(\frac{\Gamma\qty(\frac{5}{7})}
{\Gamma\qty(\frac{17}{14})})^{\frac{4}{3}}\qty(\frac{1}{h^{\frac{4}{3}}}-\frac{1}
{\qty(2\ell +h)^{\frac{4}{3}}})  \\
&\qquad\qquad +\frac{441}{16\pi}\frac{\Gamma(\frac{17}{14})^{2}}{\Gamma\qty(\frac{5}{7})^{2}}
\ell\qty(\ell +h)\qty(1-\frac{2\Gamma\qty(\frac{17}{14})\Gamma\qty(\frac{10}{7})}
{\Gamma\qty(\frac{5}{7})\Gamma\qty(\frac{27}{14})})\qty(\frac{8\pi T}{15})^{\frac{10}{3}}
+\cdots\Bigg].
\end{split}
\end{align}

On the other hand, at high temperature the corresponding EWCS behaves as
\begin{align}\label{EwD2::HT}
\begin{split}
\text{E}_{W_{HT}}^{(D2)}\approx\frac{L\qty(g_{YM}^{2}N)^{-\frac{1}{3}}}{4G_{\mathsf{N}}^{4}
\Delta^{\frac{1}{3}}}\; T^{\frac{4}{3}} & \left[\tilde{\mathcal{C}}_{2}\qty(\frac{8\pi}
{15})^{\frac{4}{3}}+\frac{3\qty(2\pi)^{\frac{2}{3}}}{21^{\frac{4}{3}}}
\qty(\frac{\Gamma\qty(\frac{5}{7})}{\Gamma\qty(\frac{17}{14})})^{\frac{4}{3}}
\qty(\frac{1}{hT})^{\frac{4}{3}}  \right.
\\
&\left. +\frac{441}{32\pi}\frac{\Gamma\qty(\frac{17}{14})^{3}\Gamma\qty(\frac{10}{7})}
{\Gamma\qty(\frac{5}{7})^{3}\Gamma\qty(\frac{27}{14})}\qty(\frac{8\pi}{15})^{\frac{10}{3}}
\qty(hT)^{2}\right],
\end{split}
\end{align}
with $\displaystyle \tilde{\mathcal{C}}_{2}=\sum_{n=0}^{\infty}\frac{3\Gamma\qty(n+\frac{1}{2})}{\sqrt{\pi}\qty(10n-4)
\Gamma \qty(n+1)}$.


\begin{flushleft}
\textbf{\underline{D3-brane}} \cite{BabaeiVelni:2019pkw,Jokela:2019ebz,Chakrabortty:2020ptb}
\end{flushleft}

The low temperature behaviour of EWCS for the D3-branes (\ref{ewcs::expn}) can be expressed as
\begin{align}\label{EwD3::LT}
\begin{split}
\text{E}_{W_{LT}}^{(D3)}&=\frac{L^{2}}{4G_{\mathsf{N}}^{5}}\sum_{n=0}^{\infty}
\frac{\Gamma\qty(n+\frac{1}{2})}{2\sqrt{\pi}\Gamma\qty(n+1)}
\qty(\frac{z_{t}\qty(2\ell+h)^{2\qty(2n-1)}-z_{t}(h)^{2\qty(2n-1)}}{2n-1})
\cdot \frac{1}{z_{0}^{4n}}  \\
&=\frac{L^{2}}{4G_{\mathsf{N}}^{5}}\Bigg[\frac{2\pi \Gamma\qty(\frac{2}{3})^{2}}
{\Gamma\qty(\frac{1}{6})^{2}}\qty(\frac{1}{h^{2}}-\frac{1}{\qty(2\ell +h)^{2}})  \\
&\qquad\qquad +\frac{9\Gamma(\frac{7}{6})^{2}}{\pi\Gamma\qty(\frac{2}{3})^{2}}
\ell\qty(\ell +h)\qty(1-\frac{2\Gamma\qty(\frac{7}{6})\Gamma\qty(\frac{4}{3})}
{\Gamma\qty(\frac{2}{3})\Gamma\qty(\frac{11}{6})})\qty(\pi T)^{4}
+\cdots\Bigg].
\end{split}
\end{align}

On the other hand, at high temperature the corresponding EWCS behaves as
\begin{align}\label{EwD3::HT}
\begin{split}
\text{E}_{W_{HT}}^{(D3)}\approx\frac{L^{2}}{4G_{\mathsf{N}}^{5}}\; T^{2} &
\left[\tilde{\mathcal{C}}_{3} \pi^{2}+\frac{4\pi\Gamma\qty(\frac{2}{3})^{2}}
{\Gamma\qty(\frac{1}{6})^{2}}\qty(\frac{1}{hT})^{2} 
+\frac{9\pi^{3}\Gamma\qty(\frac{7}{6})^{3}\Gamma\qty(\frac{4}{3})}
{\Gamma\qty(\frac{2}{3})^{3}\Gamma\qty(\frac{11}{6})}\qty(hT)^{2}\right],
\end{split}
\end{align}
with $\displaystyle \tilde{\mathcal{C}}_{3}=\sum_{n=0}^{\infty}\frac{\Gamma\qty(n+\frac{1}{2})}{2\sqrt{\pi}\qty(2n-1)
\Gamma \qty(n+1)}$.

\begin{flushleft}
\textbf{\underline{D4-brane}}
\end{flushleft}

Similar to previous three cases, the low temperature behaviour of EWCS (\ref{ewcs::expn}) corresponding to the D4-brane
is given by
\begin{align}\label{EwD4::LT}
\begin{split}
\text{E}_{W_{LT}}^{(D4)}&=\frac{2L^{3}\qty(g_{YM}^{2}N)}{G_{\mathsf{N}}^{6}}\sum_{n=0}^{\infty}
\frac{\Gamma\qty(n+\frac{1}{2})}{\sqrt{\pi}\Gamma\qty(n+1)}\qty(\frac{z_{t}
\qty(2\ell+h)^{6n-4}-z_{t}(h)^{6n-4}}{6n-4})\cdot \frac{1}{z_{0}^{6n}}  \\
&=\frac{2L^{3}\qty(g_{YM}^{2}N)}{G_{\mathsf{N}}^{6}}\Bigg[ \frac{1028 \pi^{2}
\Gamma\qty(\frac{3}{5})^{4}}{\Gamma\qty(\frac{1}{10})^{4}}\qty(\frac{1}{h^{4}}-
\frac{1}{\qty(2\ell + h)^{4}})   \\
&\qquad+\frac{25}{16\pi}\frac{\Gamma\qty(\frac{11}{10})^{2}}{\Gamma\qty(\frac{3}{5})^{2}}
\qty(1-\frac{2\Gamma\qty(\frac{11}{10})\Gamma\qty(\frac{6}{5})}{\Gamma
\qty(\frac{3}{5})\Gamma\qty(\frac{17}{10})})\ell\qty(\ell + h)\qty(\frac{8\pi T}{3})^{6}
+\cdots \Bigg].
\end{split}
\end{align}

Also, at high temperature the corresponding EWCS behaves as

\begin{align}\label{EwD4::HT}
\begin{split}
\text{E}_{W_{HT}}^{(D4)}\approx\frac{L^{3}\qty(g_{YM}^{2}N)}{4G_{\mathsf{N}}^{6}
\Delta^{3}}\;T^{4}\Bigg[ \tilde{\mathcal{C}}_{4}\qty(\frac{8\pi}{3})^{4}+\frac{1024\pi^{2}\Gamma
\qty(\frac{3}{5})^{4}}{\Gamma\qty(\frac{1}{10})^{4}}\qty(\frac{1}{hT})^{4}+\frac{25}{32\pi}
\frac{\Gamma\qty(\frac{11}{10})^{3}\Gamma\qty(\frac{6}{5})}{\Gamma\qty(\frac{3}{5})^{3}
\Gamma\qty(\frac{17}{10})}\qty(\frac{8\pi}{3})^{6}\qty(hT)^{2}\Bigg],
\end{split}
\end{align}
where $\displaystyle \tilde{\mathcal{C}}_{4}=\sum_{n=0}^{\infty}\frac{\Gamma\qty(n+\frac{1}{2})}{\sqrt{\pi} \qty(6n-4)
\Gamma\qty(n+1)}$. \hfill \break

\begin{figure}[t!]
    \centering
    {{\includegraphics[width=8.2cm]{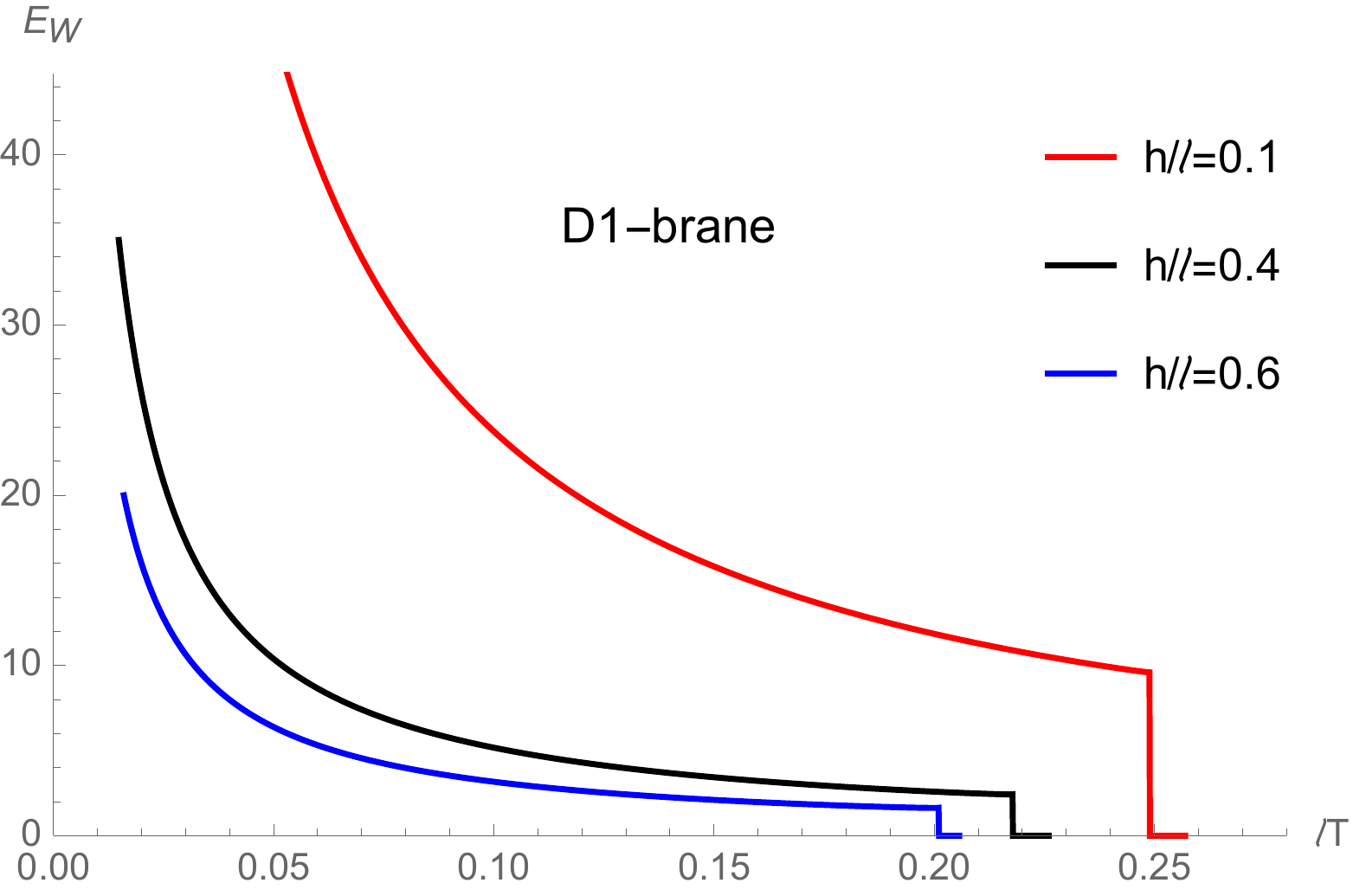} }}%
    \qquad
    {{\includegraphics[width=8.2cm]{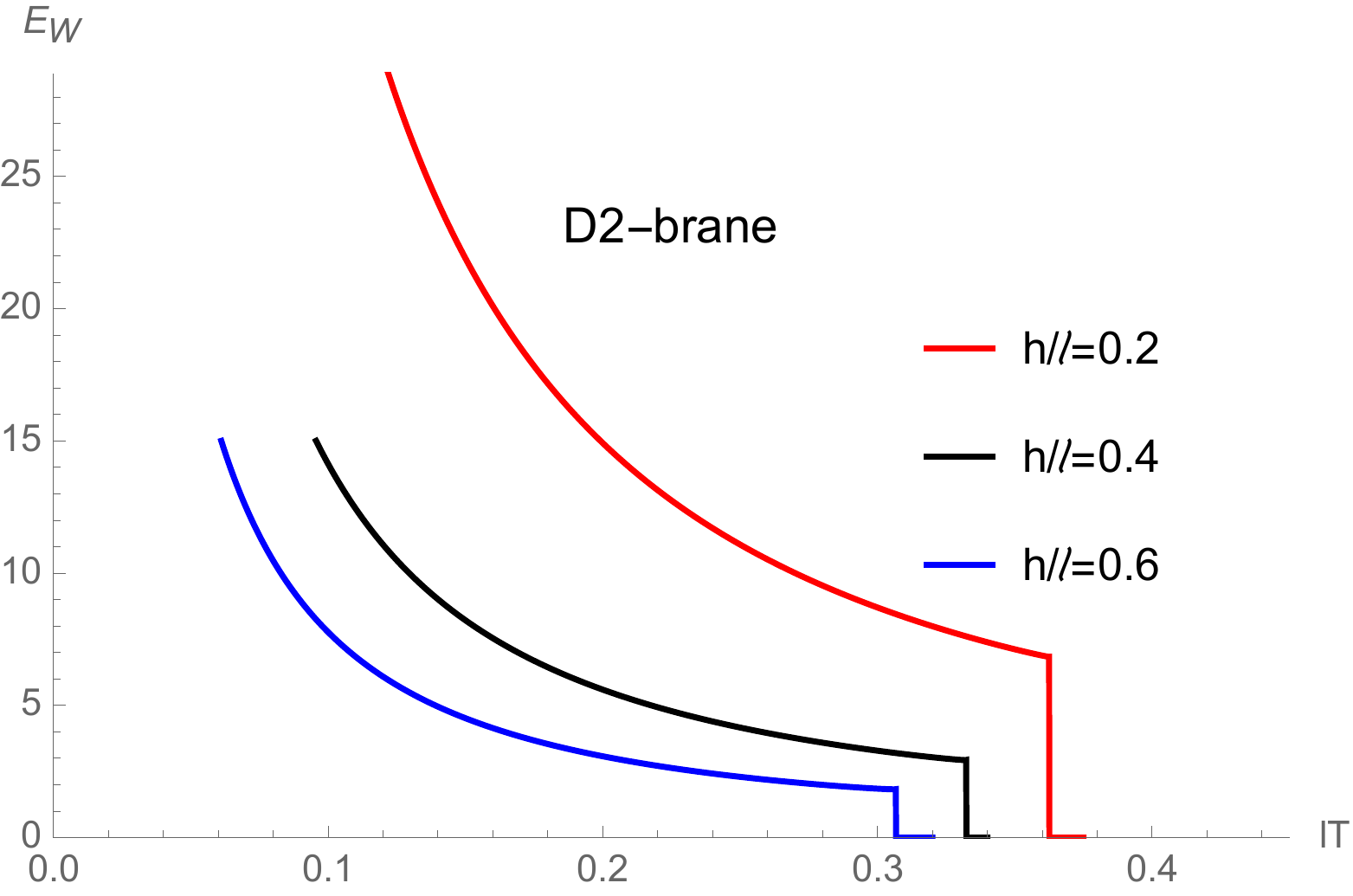} }}%
    \qquad 
   {{\includegraphics[width=8.2cm]{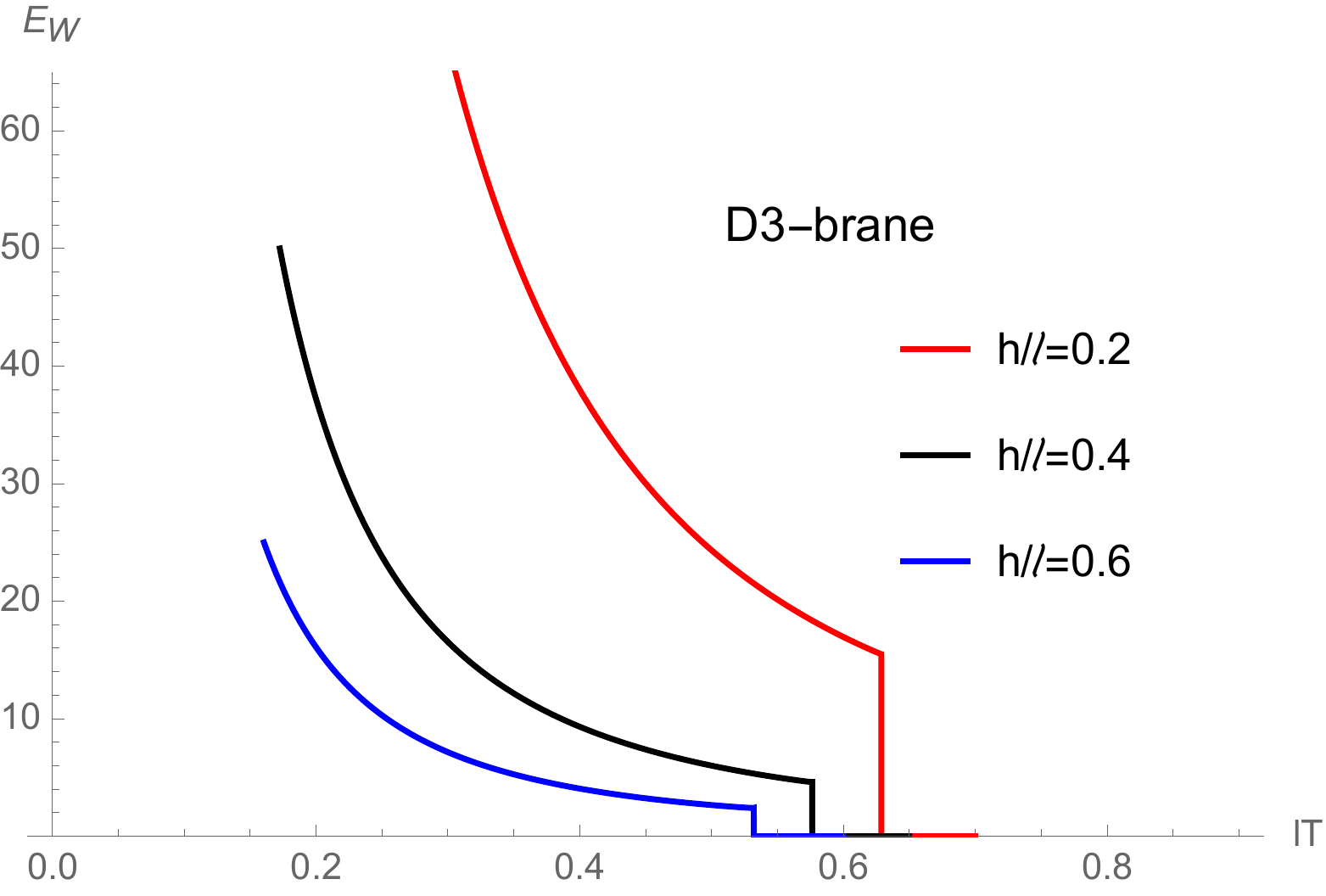} }}
   \qquad 
   {{\includegraphics[width=8.2cm]{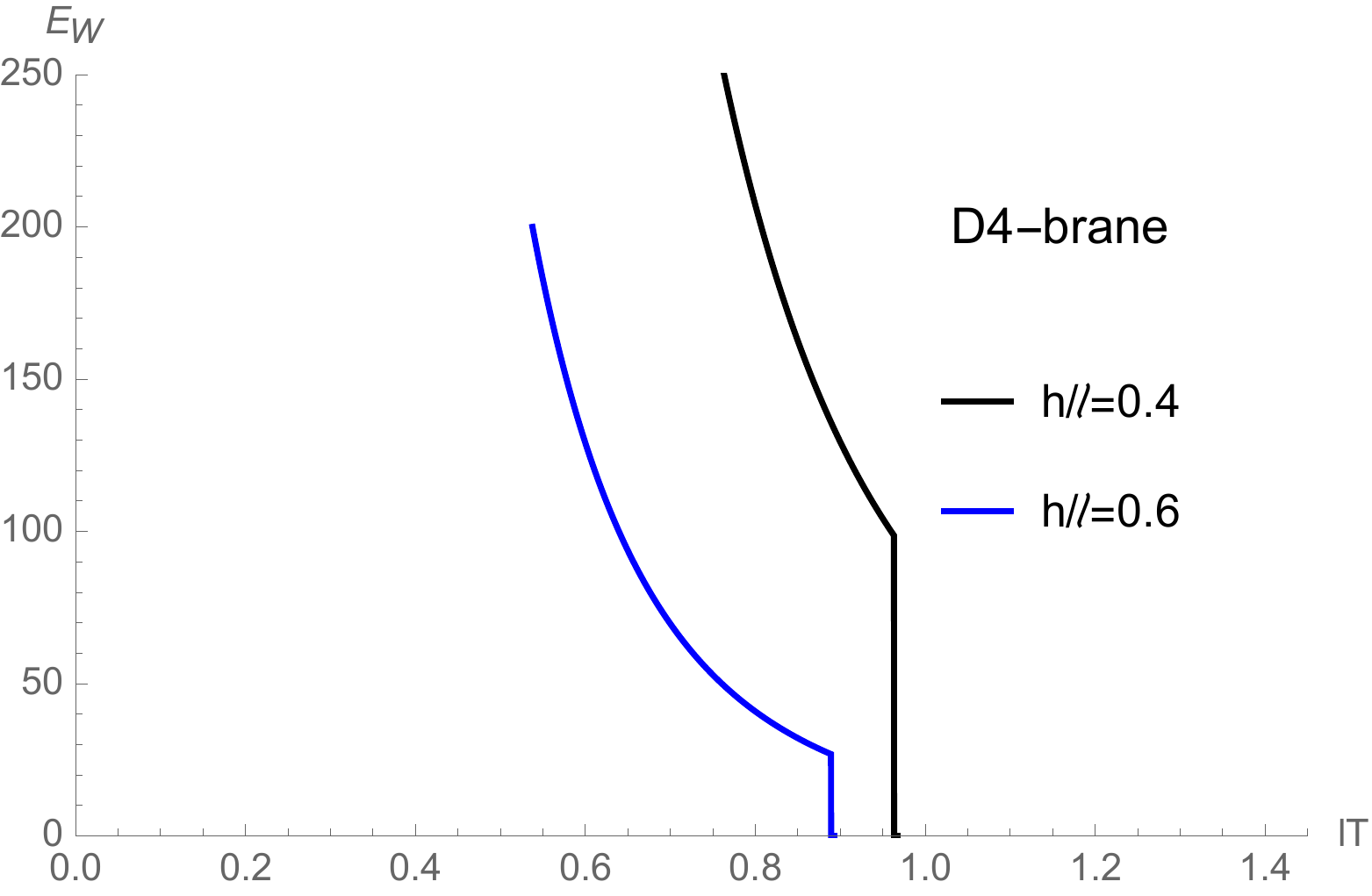} }}
\caption{$\text{E}_{W}$ \emph{vs.} $\ell T$ plots for different
values of the dimensionless quantity $h/\ell$. We have set
the pre factor $\frac{L^{p-1}}{4G_{\mathsf{N}}^{p+2}}\frac{\qty(g_{YM}^{2}N)^{\frac{
(p-3)}{(5-p)}}}{\Delta^{\frac{p-1}{5-p}}}=1$. Clearly, as the temperatures
$(\ell =1)$ increase $\text{E}_{W}$ shows a decreasing behaviour. In all
cases, the $E_{W}$ sharply drops down to zero at some critical value of
$\ell$. The behaviour of $E_{W}$ corresponding to the conformal
D$3$-brane is also shown.}
\label{Fig:2}
\end{figure}

\begin{figure}[t!]
    \centering
    {{\includegraphics[width=8.2cm]{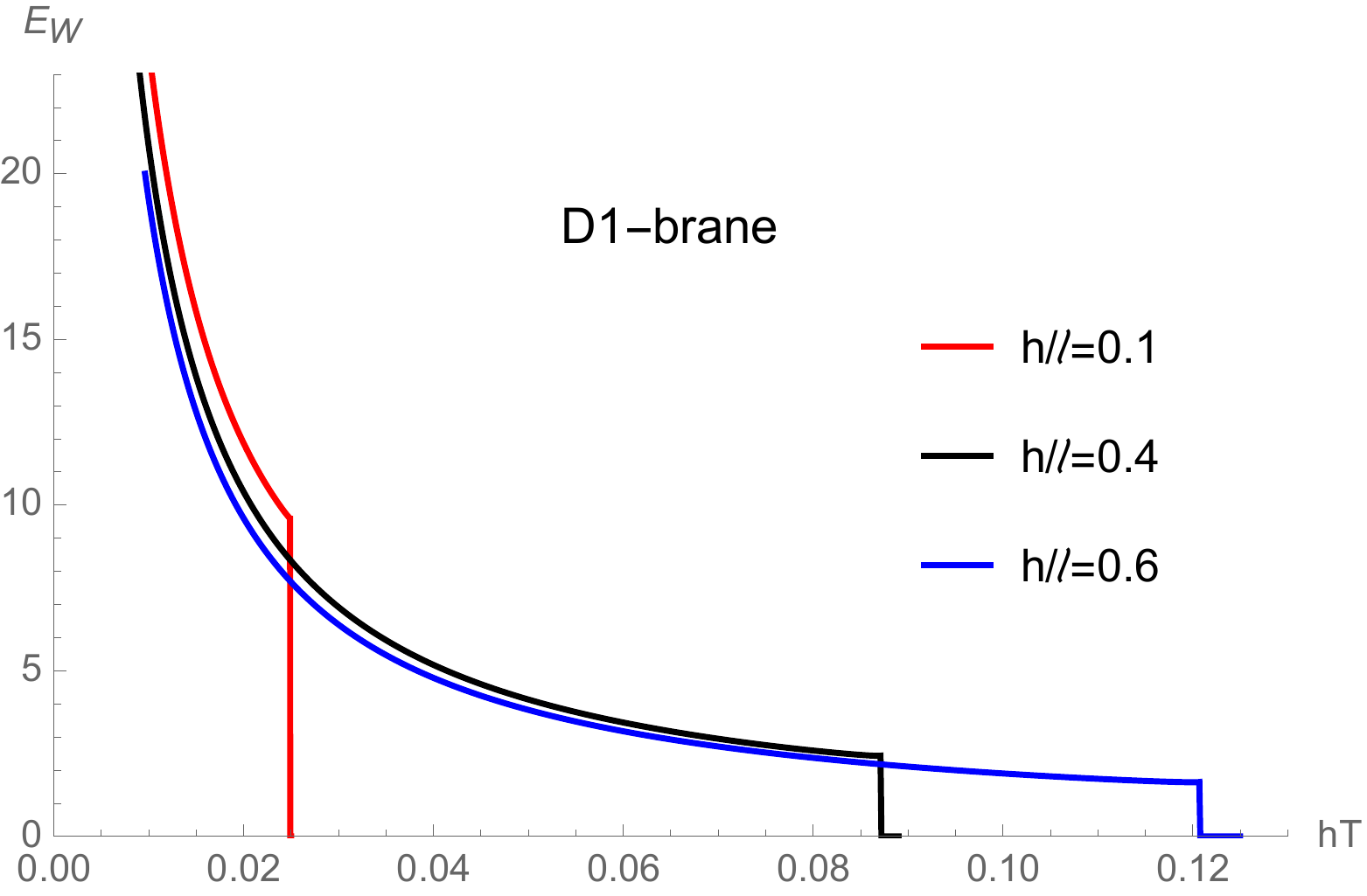} }}%
    \qquad
    {{\includegraphics[width=8.2cm]{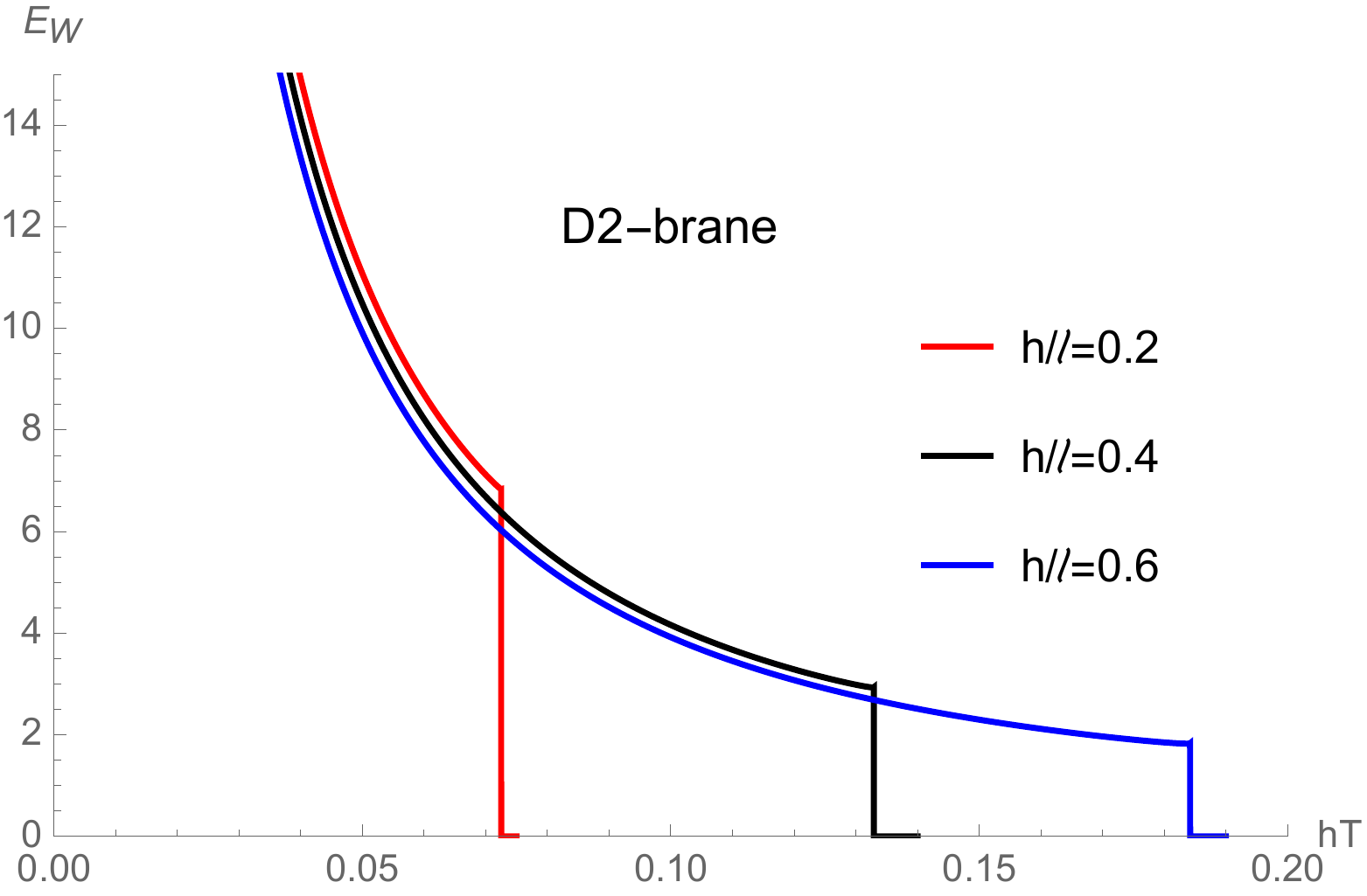} }}%
    \qquad 
   {{\includegraphics[width=8.4cm]{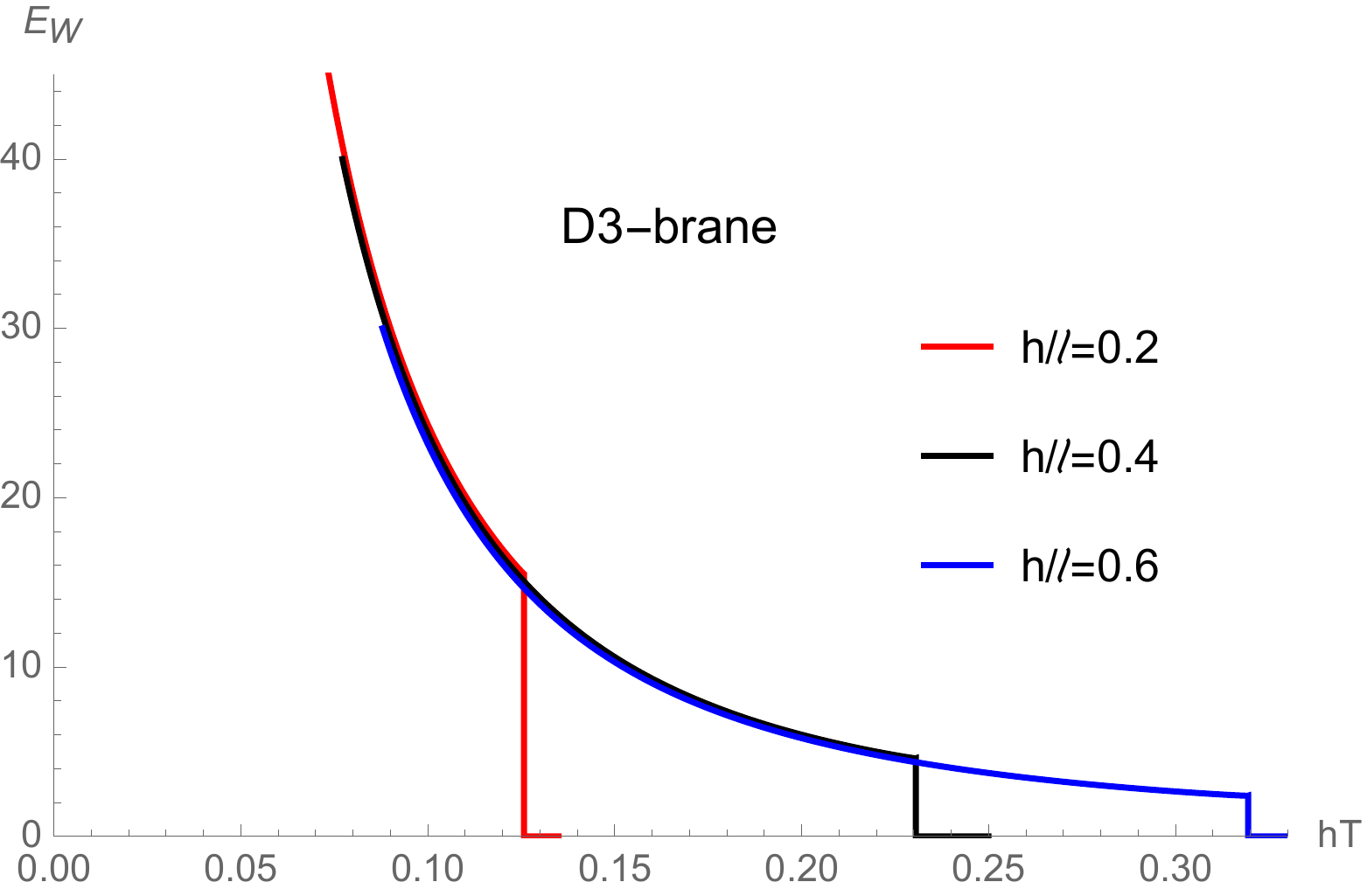} }}
   \qquad 
   {{\includegraphics[width=8.4cm]{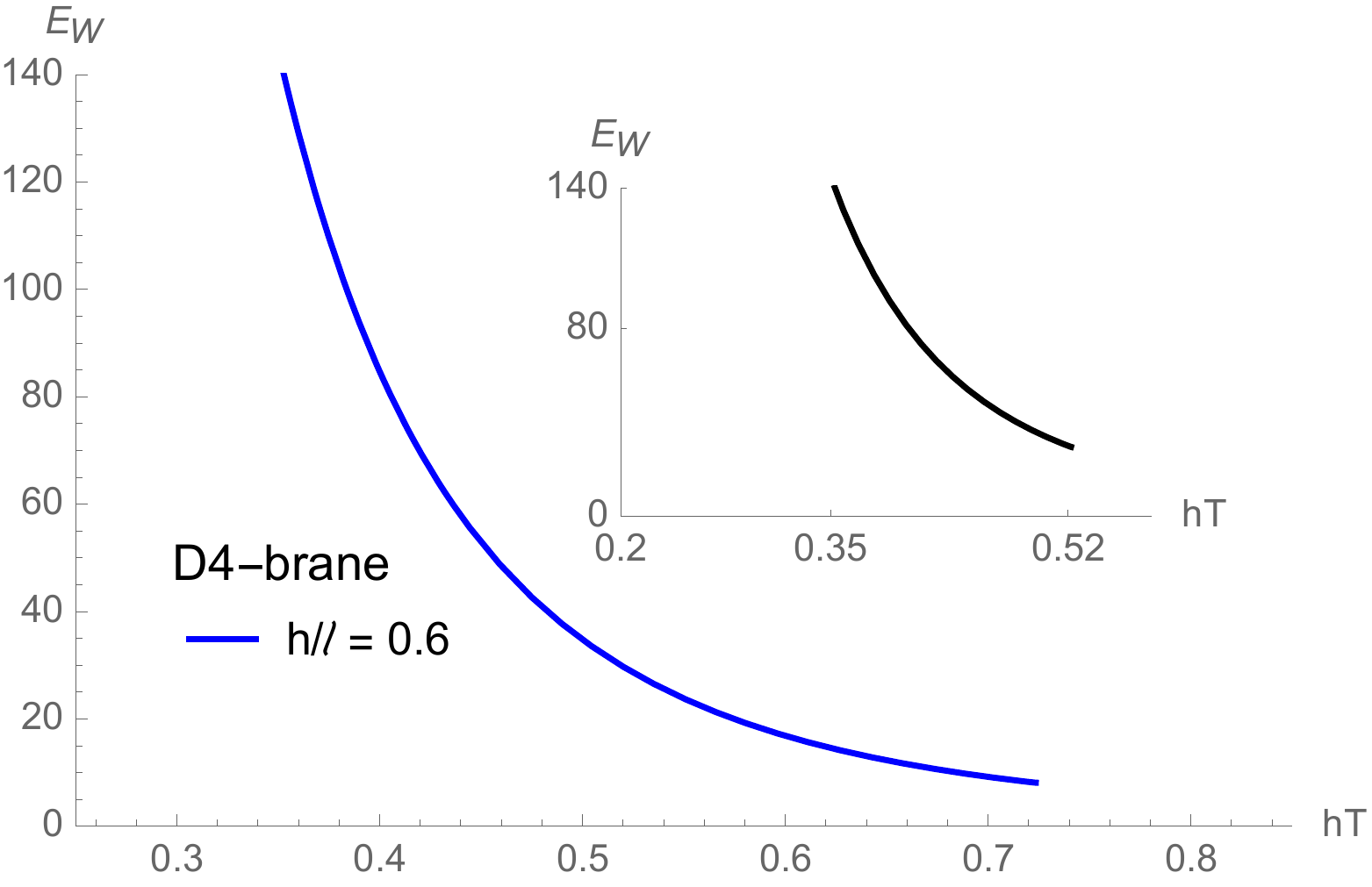} }}
\caption{$\text{E}_{W}$ \emph{vs.} $h T$ plots for different values of the dimensionless
quantity $h/\ell$. Here we have set $\frac{L^{p-1}}{4G_{\mathsf{N}}^{p+2}}\frac{\qty(g_{YM}^{2}
N)^{\frac{(p-3)}{(5-p)}}}{\Delta^{\frac{p-1}{5-p}}}=1$. For clarity, the plot corresponding to
$h/\ell=0.4$ for the D$4$-brane is given in the inset. In all cases, the $E_{W}$ jumps
discontinuously to zero at some critical value of $h$. The plot corresponding to the
conformal D$3$-brane is also given for comparison.}
\label{Fig:3}
\end{figure}

Notice that, we can numerically integrate the integration appearing in (\ref{ewcs::expn}) for given values of $p$ exactly.
In fig.(\ref{Fig:2}) and fig.(\ref{Fig:3}) we plot numerically the behaviour of $E_{W}$ for different values of the ratio $h/\ell$
for different D$p$-brane backgrounds.


Let us now discuss the following observations that we can make from the low as well as high temperature behaviours of
the EWCS derived above. These are listed as follows:
\begin{itemize}

\item When the widths of the subsystems ($\ell$) are small EWCS increase sharply (fig.\ref{Fig:2}). On the
other hand, for a given value of the ratio $h/\ell$, it converges to a finite value as $\ell$ increases. Also, beyond
certain values of the width it sharply drops down to zero value indicating a phase transition due to the
disentanglement of the subsystems. For the intermediate values of $\ell$, the decrease in EWCS is monotonic.

\item At low temperatures, the first terms of the EWCS in (\ref{EwD1::LT}), (\ref{EwD2::LT}),
(\ref{EwD3::LT}) and (\ref{EwD4::LT}) increase as the separation $h$ between the two entangling regions
decrease. Moreover, as $h \to 0$ the EWCS diverge. This is clearly visible from fig.\ref{Fig:3}. Also, if we
keep the width and separation between the two subregions fixed and increase the temperature, the EWCS
decreases monotonically to zero ($ \text{E}_{W} = 0 $) in all cases indicating a phase transition resulting
due to the disentanglement of the Ryu-Takayanagi surfaces.

\item The first terms in (\ref{EwD1::HT}), (\ref{EwD2::HT}), (\ref{EwD3::HT}) and (\ref{EwD4::HT}) is proportional
to the area of the entangling region, $L^{p-1}$. This suggests that at  finite temperature the EWCS obeys an
area law. This is in sharp contrast to the HEE at high temperature obtained in (\ref{HEE::HT}) where it scales
with the volume. Similar observations have in fact been made earlier \cite{BabaeiVelni:2019pkw,Jokela:2019ebz, 
Chakrabortty:2020ptb}. Moreover, this area law scaling implies that the EWCS carries more information than the
HEE regarding the correlation between $A$ and $B$ \cite{Wolf:2007tdq,Headrick:2010zt,Fischler:2012uv,
BabaeiVelni:2019pkw,Jokela:2019ebz,Chakrabortty:2020ptb}.

\end{itemize}

\subsection{Critical separation between the strips}\label{separation}
In this section, we determine the critical separation between the parallel strips $A$ and $B$ at which these two subregions
become completely disentangled. This can indeed be found by calculating the holographic mutual information (HMI) between
$A$ and $B$ \cite{Ben-Ami:2014gsa,Wolf:2007tdq,Headrick:2010zt,Fischler:2012uv}. The HMI between the two subregions
$A$ and $B$ can be expressed as \cite{Ben-Ami:2014gsa,Wolf:2007tdq,Headrick:2010zt,Fischler:2012uv,
BabaeiVelni:2019pkw,Jokela:2019ebz,Chakrabortty:2020ptb}
\begin{equation}\label{HMI}
\mathtt{I}_{M}\qty(\ell,h)=2S\qty(\ell) - S\qty(h) - S\qty(2\ell + h).
\end{equation}
It measures the total correlations between $A$ and $B$.

At the critical value of the separation, $h_{c}$, at which the two subregions are no longer entangled, we have
\begin{equation}\label{cricl::HMI}
\texttt{I}_{M}\qty(\ell,h_{c})=0.
\end{equation}  

This is reminiscent of a first order phase transition which occurs due to the competition between different configurations
for calculating $S\qty(2\ell + h)$: when the separation distance is small the connected configuration ($S_{\text{con}}$)
is preferred over the disconnected ($S_{\text{dis}}$) one. On the other hand, for large separation the disconnected
configuration is preferred resulting the vanishing of the HMI \cite{Ben-Ami:2014gsa,Headrick:2010zt,BabaeiVelni:2019pkw}.
This type of phase transitions occur for large values of the subsystem size $\ell \to \infty$. Thus $S(\ell)$ and $S \qty
(2\ell+h)$ are represented by the IR (high temperature) expression for the HEE while $S(h)$ is represented by the UV (low 
temperature) expression for the HEE 
\cite{Takayanagi:2017knl,Nguyen:2017yqw,BabaeiVelni:2019pkw,Jokela:2019ebz,Chakrabortty:2020ptb}. 

Before proceeding to the finite temperature analysis, we notice that in the limit $\ell T\to 0$, i.e., very close to zero
temperature, one can calculate the critical values of the ratio $\chi = h/\ell$ from the leading finite terms of (\ref{HEED1::LT}), 
(\ref{HEED2::LT}), (\ref{HEED3::LT}) and (\ref{HEED4::LT}). In order to do that, we note that in the disconnected phase the 
expression for HEE may be written as \cite{Ben-Ami:2014gsa}
\begin{equation}\label{hee::dis}
S_{\text{dis}}=2S\qty(\ell),
\end{equation}
whereas, for the connected phase
\begin{equation}\label{hee::con}
S_{\text{con}}=S\qty(h)+S\qty(2\ell +h).
\end{equation}

Now, along the transition lines $\chi_{c}=\text{constant}$, $S_{\text{dis}}=S_{\text{con}}$. This allows us to write the
corresponding equations for $\chi_{c}$ as \cite{Ben-Ami:2014gsa}
\begin{equation}\label{hl::eqn}
2=\begin{cases}
               \frac{1}{\chi_{c}}+\frac{1}{\qty(2+\chi_{c})} & \text{for D1-brane,}\\[10pt]
               \frac{1}{\chi_{c}^{4/3}}+\frac{1}{\qty(2+\chi_{c})^{4/3}} & \text{for D2-brane,}\\[10pt]
               \frac{1}{\chi_{c}^{2}}+\frac{1}{\qty(2+\chi_{c})^{2}} & \text{for D3-brane,}\\[10pt]
               \frac{1}{\chi_{c}^{4}}+\frac{1}{\qty(2+\chi_{c})^{4}} & \text{for D4-brane.}
            \end{cases}
\end{equation}

From (\ref{hl::eqn}) we can compute the critical values as $\chi_{c}=0.618034, 0.663183, 0.732051, 0.842514$ for D$1$-,
D$2$-, D$3$-, D$4$-brane, respectively. We also observe that, as $p$ increases, the slope of the corresponding critical
line also increases.

\begin{figure}[t!]
    \centering
    {{\includegraphics[width=8.2cm]{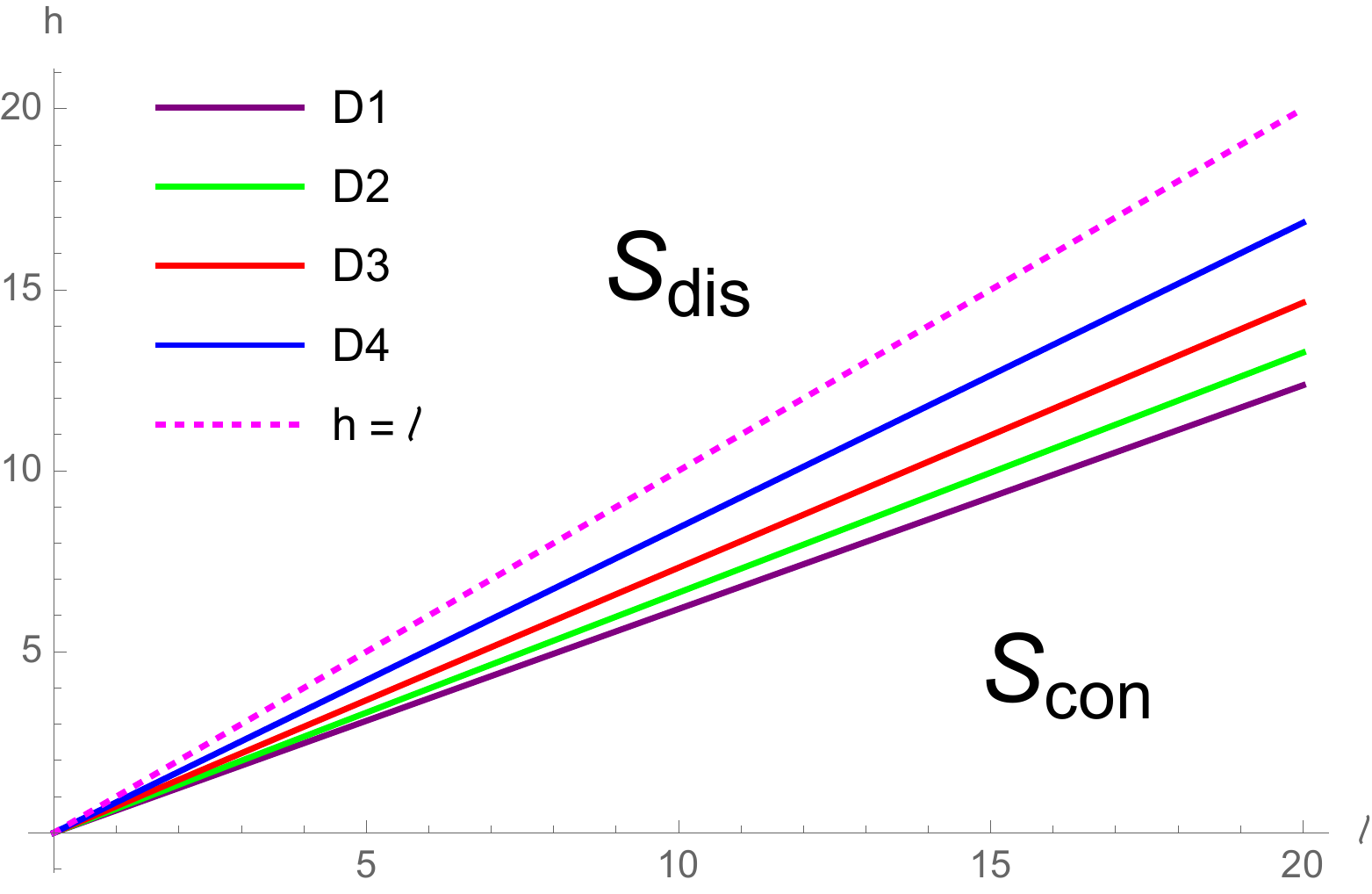} }}%
    \qquad
    {{\includegraphics[width=8.2cm]{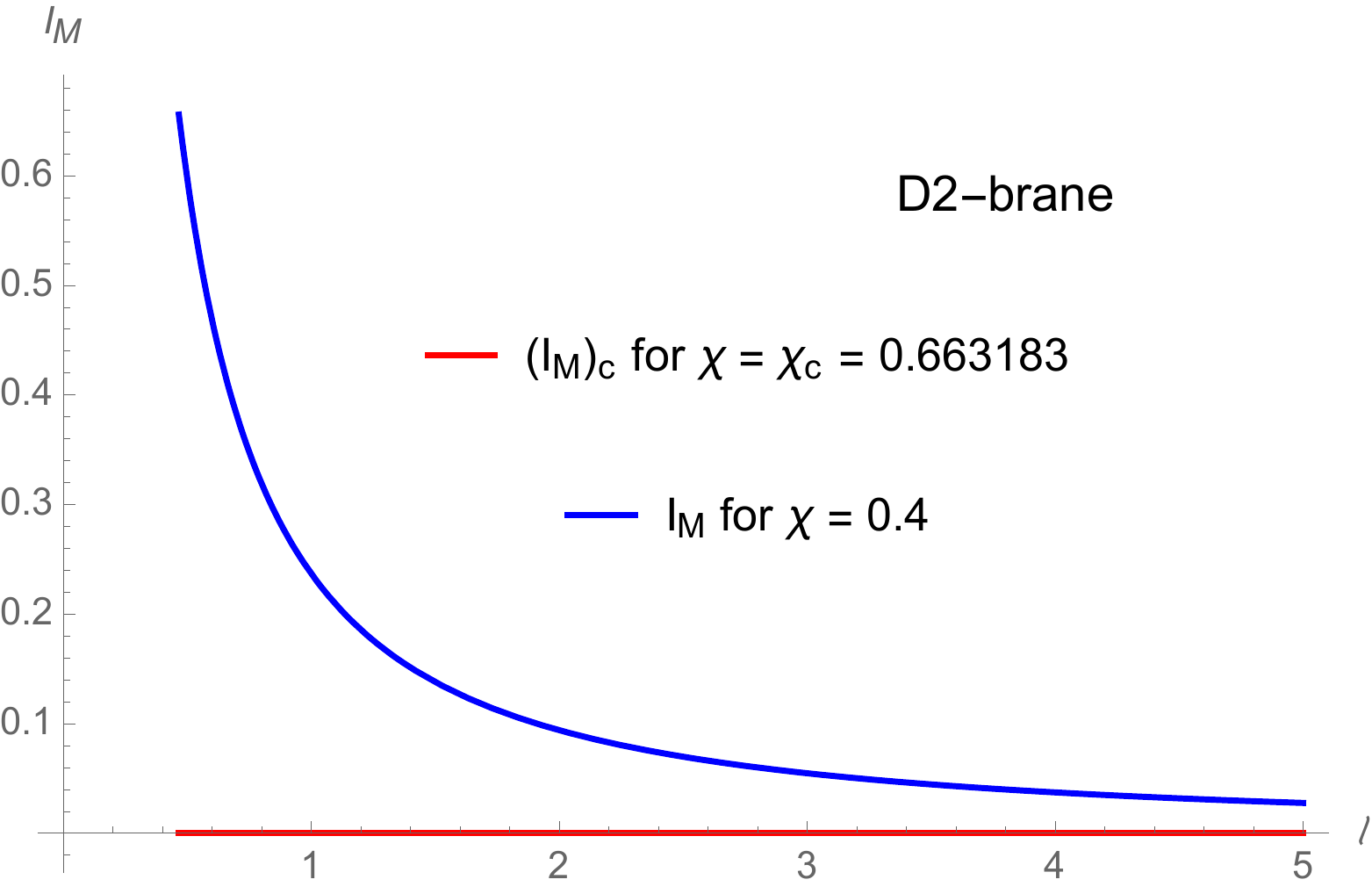} }}%
\caption{\textbf{A:} The phase diagrams corresponding to different
D$p$-branes. The entire region above an individual curve corresponds to
the disconnected phase ($\mathbf{S_{\text{\textbf{dis}}}}$), while that
below an individual curve corresponds to the connected phase
($\mathbf{S_{\text{\textbf{con}}}}$). The dotted line corresponds to
$h=\ell$. \textbf{B:} Mutual information ($\texttt{I}_{M}$) \textit{versus}
strip width ($\ell$) plot for D$2$-brane.}
\label{Fig:phase}
\end{figure}

In fig.\ref{Fig:phase}A, we have shown the phase diagrams corresponding to the D$p$-branes. Whereas, in
fig.\ref{Fig:phase}B the behaviour of the mutual information ($\texttt{I}_{M}$) has been shown for D$2$-brane. Similar plots
can be drawn for other backgrounds as well. Notice that, as long as $\chi < \chi_{c}$ ($\chi_{c}=0.663183$ for D$2$-brane 
\cite{Ben-Ami:2014gsa}), $\texttt{I}_{M}$ is a monotonically decreasing function of $\ell$; while at $\chi_{c}$ it vanishes.
Similar bahaviours of $\texttt{I}_{M}$ were observed for pure AdS$_{d+1}$ backgrounds in \cite{Jokela:2019ebz}. Also
note that, in plotting fig.\ref{Fig:phase}B numerically we have used (\ref{HEE:new}) and (\ref{HMI}).

We now move on to the finite temperature cases and compute the critical separation, $h_{c}$. However, due to the
complexity of the resulting expressions that arise from (\ref{cricl::HMI}), below we only write down the general form
of the equation for the critical separations ($h_{c}$) as
\begin{equation}\label{eqn::cric}
\sum_{i=0}^{4}\text{a}_{i}\qty(\frac{h_{c}}{z_{0}})^{\text{n}_{i}}=0,
\end{equation}
where we have used results from section \ref{LT::HEE}. In (\ref{eqn::cric}), the numerical coefficients $\text{a}_{i}$ are
certain combinations of gamma functions whose explicit expressions can be written in terms of various constants that
appear in section \ref{LT::HEE} and $\Theta$ in (\ref{HEE::HT}) and are presented in table \ref{Tab::A} in appendix
\ref{coefs}. On the other hand, the exponents $\text{n}_{i}$ may be given as\footnote{The corresponding expression for
D$3$-brane may be compared with that already available in the literature; see, e.g., \cite{Jokela:2019ebz,
Chakrabortty:2020ptb}.}
\begin{align}
\begin{split}
\text{for D1-brane:} & \qquad\qquad \qty(\text{n}_{0}=0,\text{n}_{1}=1,
\text{n}_{2}=2,\text{n}_{3}=3,\text{n}_{4}=6),  \\[7pt]
\text{for D2-brane:} & \qquad\qquad \qty(\text{n}_{0}=0,\text{n}_{1}=\frac{4}{3},
\text{n}_{2}=\frac{7}{3},\text{n}_{3}=\frac{10}{3},\text{n}_{4}=\frac{20}{3}),  \\[7pt]
\text{for D3-brane:} & \qquad\qquad \qty(\text{n}_{0}=0,\text{n}_{1}=2,
\text{n}_{2}=3,\text{n}_{3}=4,\text{n}_{4}=8),  \\[7pt]
\text{for D4-brane:} & \qquad\qquad \qty(\text{n}_{0}=0,\text{n}_{1}=4,
\text{n}_{2}=5,\text{n}_{3}=6,\text{n}_{4}=12). 
\end{split}
\end{align}

\begin{figure}[t!]
    \centering
    {{\includegraphics[width=8.2cm]{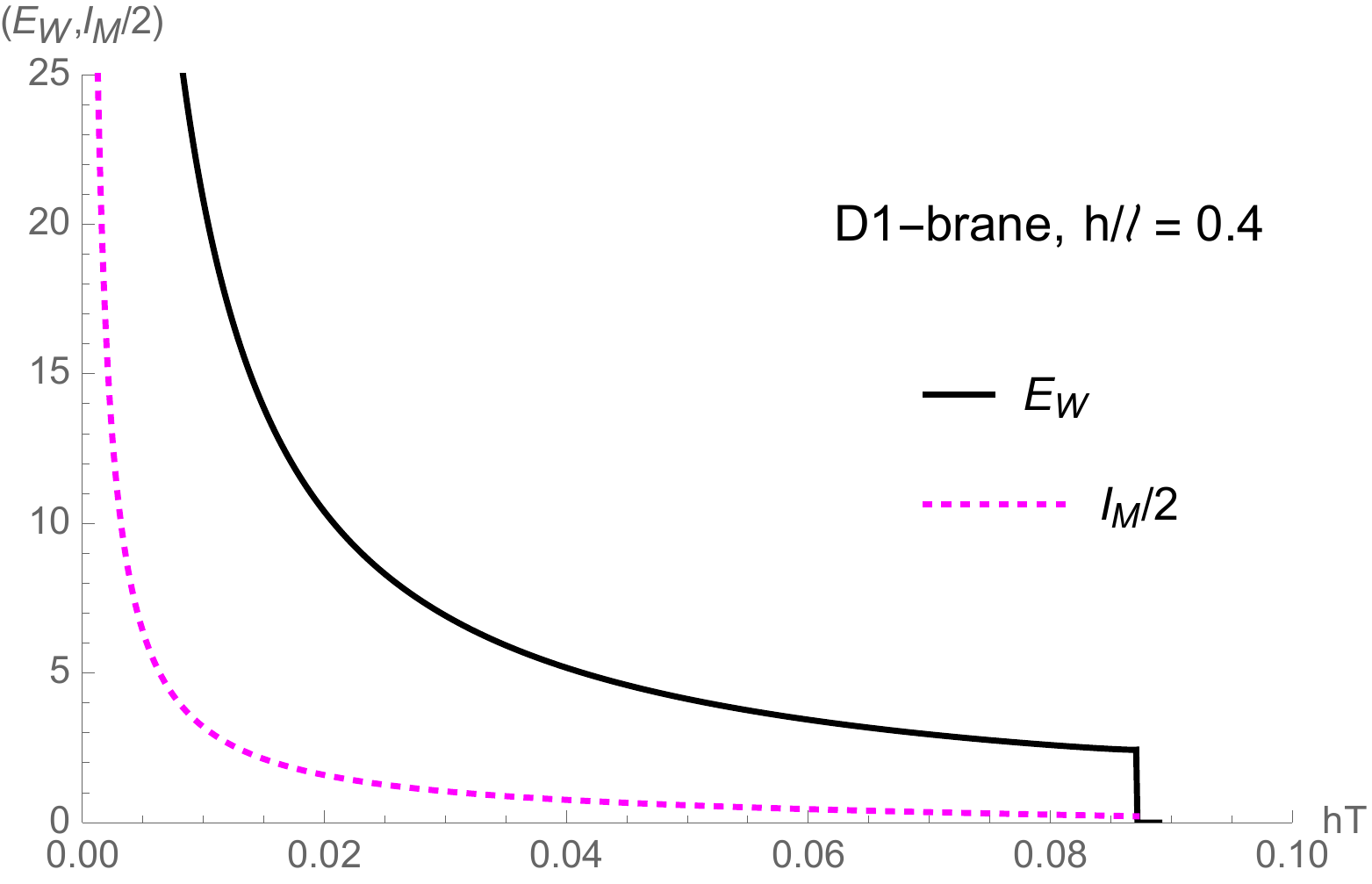} }}%
    \qquad
    {{\includegraphics[width=8.2cm]{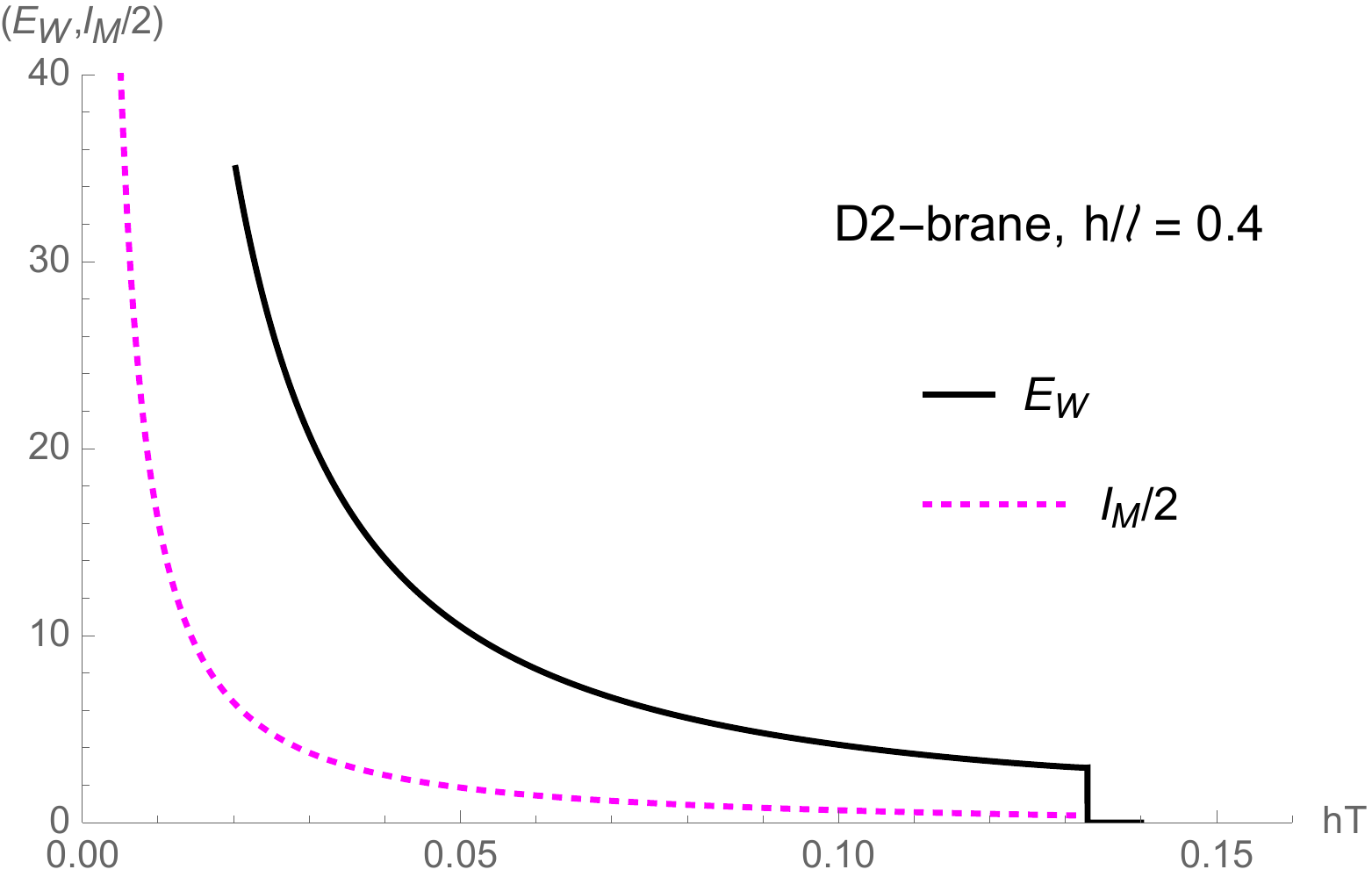} }}%
    \qquad 
  {{\includegraphics[width=8.4cm]{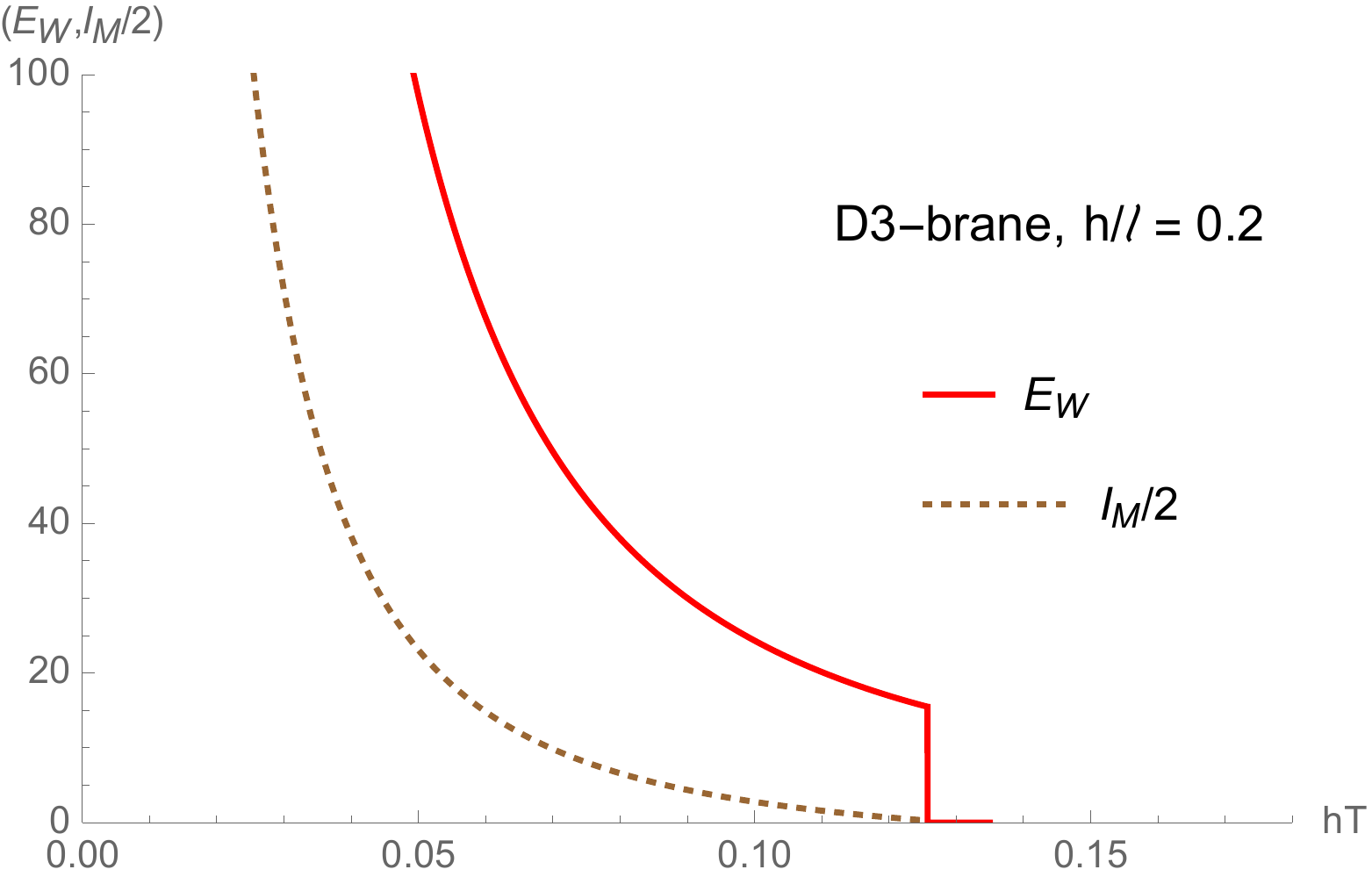} }}
  \qquad 
  {{\includegraphics[width=8.4cm]{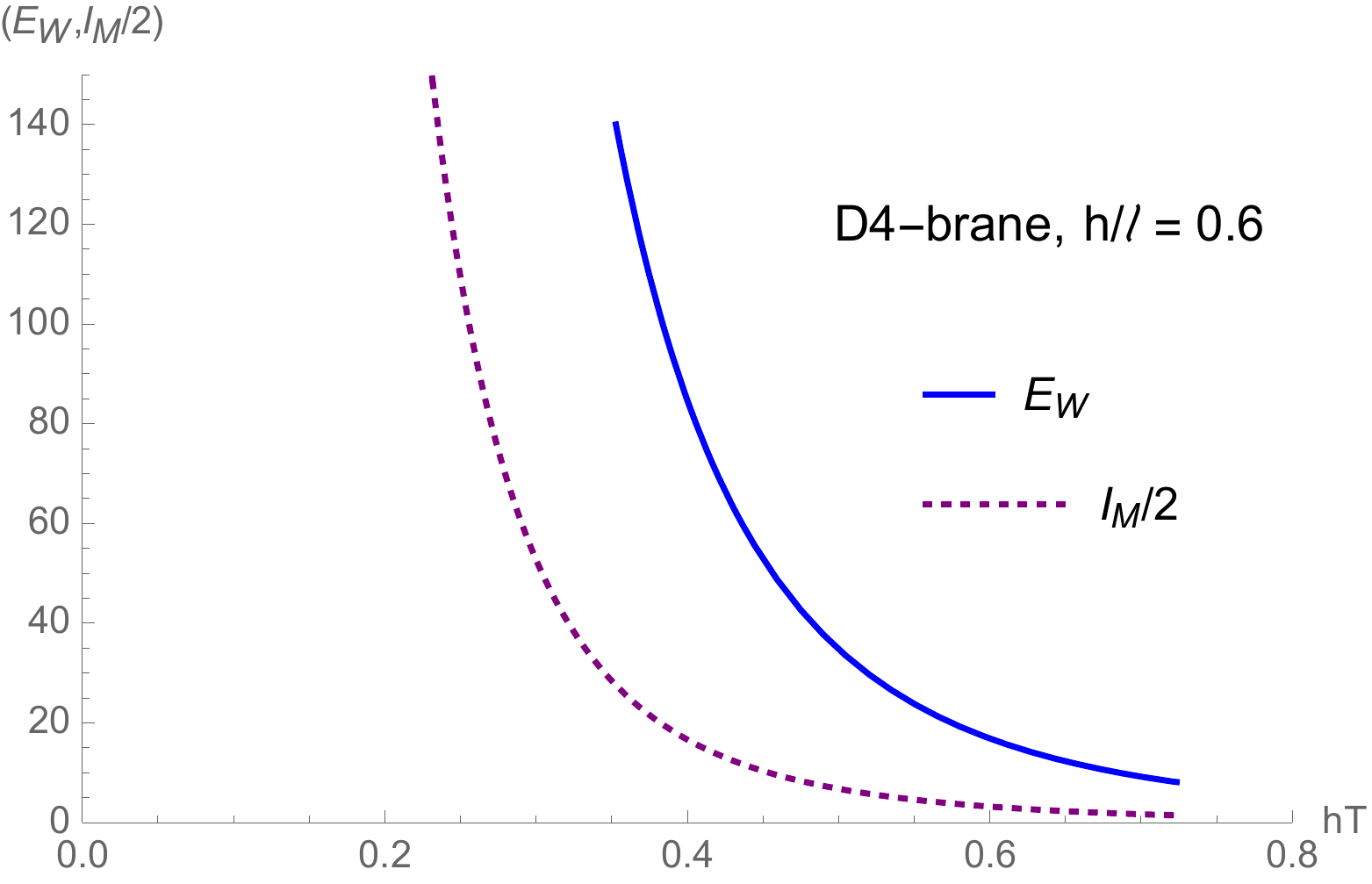} }}
\caption{Holographic Mutual Information ($\texttt{I}_{M}/2$) plots for different
non-conformal backgrounds. Notice that, in all the cases $\text{E}_{W} \geq
\texttt{I}_{M}/2$. We set $\frac{L^{p-1}}{4G_{\mathsf{N}}^{p+2}}\frac{\qty(g_{YM}^{2}
N)^{\frac{(p-3)}{(5-p)}}}{\Delta^{\frac{p-1}{5-p}}}=1$}
\label{Fig:4}
\end{figure}

Finally, we explicitly check the inequality (\ref{rel::EwMI}) by numerically plotting both $\text{E}_{W}$ and
$\texttt{I}_{M} \qty(A,B)/2$ against the dimensionless quantity $hT$ in fig.\ref{Fig:4}. Note that, in plotting fig.\ref{Fig:4}
we have used (\ref{HEE:new}) and (\ref{HMI}). In addition to validate (\ref{rel::EwMI}), these plots also show the
monotonically decreasing nature of both $\text{E}_{W}$ and $\texttt{I}_{M} \qty(A,B)$. Moreover, beyond the critical
separation $h_{c}T$ ($T=1$) determined by (\ref{eqn::cric}) both of them approach zero which is an essential nature
for the phase transition discussed above \cite{Ben-Ami:2014gsa,Headrick:2010zt,Takayanagi:2017knl, 
Nguyen:2017yqw,Terhal:2002bm}.

\section{Conclusions}\label{conclusion}
In this paper, we have studied aspects of holographic entanglement measures for certain class of non-conformal field
theories which are holographic dual to non-conformal D$p$-brane backgrounds
\cite{Itzhaki:1998dd,Boonstra:1998mp,Kanitscheider:2008kd,Pang:2013lpa,Pang:2015lka}. Within the framework of
the Gauge/Gravity duality, we explicitly computed entanglement wedge cross section (EWCS), holographic
entanglement entropy (HEE) and holographic mutual information (HMI) for these field theories. In our analysis, we
considered $p=1,2,4$; while $p=3$ corresponds to the usual AdS$_5$/CFT$_4$ duality \cite{Maldacena:1997re}.
In our computations, we considered a bipartite system with two parallel entangling regions of equal width $\ell$ and
length $L$ separated by a distance $h$. Implementing analytical as well as numerical methods, we observe that the
qualitative behaviours of the above mentioned holographic measures are similar to those of the conformal case: both
the EWCS and HMI show monotonically decreasing behaviour with temperature. However, while the previous one
discontinuously drops down to zero value beyond a critical separation distance $h_{c}$ between the entangling
subregions, the later attains zero value at $h_c$ in a smooth and continuous fashion. This suggests a first order phase
transition between connected and disconnected phases of the bipartite system. In our analysis, we have also been
able to compute this critical values analytically. Our analysis indeed establishes the fact that the qualitative behaviours
of the holographic information quantities are similar irrespective of whether the boundary field theory is conformal or
not.

We believe that this non trivial analysis of information quantities, in the light of Gauge/Gravity correspondence, will be
extremely useful in order to understand aspects of field theories with explicit breaking of conformal invariance. From this
point of view, it seems fascinating to consider quantum corrections to both EWCS and HEE in these holographic
set-ups \cite{Faulkner:2013ana}. On top of that, other relevant measures, such as, entanglement negativity
\cite{Kudler-Flam:2018qjo}, reflected entropy \cite{Dutta:2019gen,Jeong:2019xdr}, etc., can also be studied within these
non-conformal theories. It will be equally interesting to study the evolution and scaling of EWCS after thermal as well as 
electromagnetic quenches \cite{Yang:2018gfq,Kudler-Flam:2020url,BabaeiVelni:2020wfl} corresponding to these theories.
These we plan to explore in future.

\section*{Acknowledgments}
The present work is supported by the Chilean \emph{National Agency for Research and Development}
(ANID)/ FONDECYT / POSTDOCTORADO BECAS CHILE / Project No. 3190021. I would like to thank
Shankhadeep Chakrabortty, Suvankar Dutta, Radouane Gannouji, M.~Reza~Mohammadi~Mozaffar
and especially Karunava Sil for very useful discussions. I would also like to thank the referee for her/his
valuable comments.\\

\textbf{Dedication:} This article is dedicated to the memory of Mr. Anup Lala.


\appendix
\numberwithin{equation}{section}
\section{Various coefficients appearing in section (\ref{LT::HEE})}\label{coefs}
The coefficients appearing in (\ref{HEED1::LT}) [D$1$-brane] are given by
\begin{align}\label{app::D1}
\begin{split}
\mathcal{C}_{1}&=\frac{\pi \Gamma \left(\frac{-1}{4}\right) \Gamma \left(\frac{3}{4}\right)}
{16 \Gamma \left(\frac{1}{4}\right) \Gamma \left(\frac{5}{4}\right)},  \\
\mathcal{C}_{2}&=\frac{1}{\mathcal{C}_{1}}\frac{16 \Gamma \left(\frac{5}{4}\right)^2}{\Gamma
\left(\frac{3}{4}\right)^2}\qty(1+\frac{\Gamma \left(\frac{-1}{4}\right)\Gamma \left(\frac{5}{4}\right)}{2\Gamma
\left(\frac{1}{4}\right) \Gamma \left(\frac{3}{4}\right)}),    \\
\mathcal{C}_{3}&=\frac{1}{\mathcal{C}_{1}}\frac{4096 \Gamma \left(\frac{5}{4}\right)^6}{\pi\Gamma
\left(\frac{3}{4}\right)^6}\qty[\frac{4\Gamma\qty(\frac{3}{4})}{5\pi^{3/2}}+\frac{3\Gamma\qty(\frac{9}{4})
\Gamma\qty(\frac{-1}{4})}{2\pi \Gamma\qty(\frac{1}{4})\Gamma\qty(\frac{11}{4})}-1].
\end{split}
\end{align}

\noindent The coefficients appearing in (\ref{HEED2::LT}) [D$2$-brane] are written as
\begin{align}\label{app::D2}
\begin{split}
\mathcal{D}_{1}&=\frac{2^{8/3} \pi ^{7/6}\Gamma \left(\frac{-2}{7}\right) }{49 \sqrt[3]{21}
\;\Gamma \left(\frac{3}{14}\right)}\left(\frac{\Gamma \left(\frac{5}{7}\right)}{\Gamma
\left(\frac{17}{14}\right)}\right)^{4/3},  \\
\mathcal{D}_{2}&=\frac{1}{\mathcal{D}_{1}} \frac{63 \Gamma\qty(\frac{17}{14})^{3}}
{8\sqrt{\pi}\Gamma\qty(\frac{5}{7})^{3}}\qty[\frac{\Gamma\qty(\frac{-2}{7})\Gamma
\qty(\frac{10}{7})}{\Gamma\qty(\frac{3}{14})\Gamma\qty(\frac{27}{14})}+\frac{3
\Gamma\qty(\frac{3}{7})\Gamma\qty(\frac{5}{7})}{\Gamma\qty(\frac{17}{14})\Gamma
\qty(\frac{13}{4})}],  \\
\mathcal{D}_{3}&=\frac{1}{\mathcal{D}_{1}} \Bigg[-\frac{1750329 \sqrt[3]{21}\Gamma
\left(\frac{3}{7}\right) \Gamma \left(\frac{17}{14}\right)^{19/3} \Gamma
\left(\frac{10}{7}\right)}{2048\ 2^{2/3} \pi ^{13/6} \Gamma \left(\frac{5}{7}\right)^{19/3}
\Gamma \left(\frac{13}{14}\right) \Gamma \left(\frac{27}{14}\right)}    \\
& +\frac{2^{2/3} \pi^{7/6}\Gamma \left(\frac{-2}{7}\right)}{245 \sqrt[3]{21}\;\Gamma
\left(\frac{3}{14}\right)} \left(\frac{\Gamma\left(\frac{5}{7}\right)}{\Gamma \left(\frac{17}{14}
\right)}\right)^{4/3}\Bigg( \frac{333534915\ 21^{2/3} \Gamma \left(\frac{17}{14}\right)^{26/3}
\Gamma\left(\frac{10}{7}\right)^2}{4096 \sqrt[3]{2} \pi ^{10/3} \Gamma \left(\frac{5}{7}\right)^{26/3}
\Gamma \left(\frac{27}{14}\right)^2}    \\
& + \frac{142943535\ 21^{2/3}\Gamma \left(\frac{17}{14}\right)^{20/3} \left(3 \Gamma
\left(\frac{5}{7}\right) \Gamma \left(\frac{17}{14}\right) \Gamma \left(\frac{27}{14}\right)^2 \Gamma
\left(\frac{15}{7}\right)-2 \Gamma \left(\frac{17}{14}\right)^2 \Gamma \left(\frac{10}{7}\right)^2 \Gamma
\left(\frac{37}{14}\right)\right)}{4096 \sqrt[3]{2} \pi ^{10/3} \Gamma \left(\frac{5}{7}\right)^{26/3} \Gamma
\left(\frac{27}{14}\right)^2 \Gamma \left(\frac{37}{14}\right)} \Bigg) +   \\
& \frac{5250987 \sqrt[3]{21} \Gamma \left(\frac{8}{7}\right) \Gamma
\left(\frac{17}{14}\right)^{16/3}}{8192\ 2^{2/3} \pi ^{13/6} \Gamma \left(\frac{5}{7}\right)^{16/3} \Gamma
\left(\frac{23}{7}\right)}\Bigg].
\end{split}
\end{align}

\noindent The coefficients in (\ref{HEED3::LT}) [D$3$-brane] are of the following forms:
\begin{eqnarray}\label{app::D3}
\mathcal{B}_{1}&=& \frac{\pi ^{3/2}\Gamma \left(\frac{-1}{3}\right) \Gamma \left(\frac{2}{3}\right)^2}
{27 \;\Gamma \left(\frac{1}{6}\right) \Gamma \left(\frac{7}{6}\right)^2}, \nonumber \\ 
\mathcal{B}_{2}&=&\frac{1}{\mathcal{B}_{1}} \qty(\frac{3\Gamma \left(\frac{4}{3}\right) \Gamma
\left(-\frac{1}{3}\right) \Gamma \left(\frac{7}{6}\right)^3}{\sqrt{\pi }\Gamma \left(\frac{1}{6}\right)
\Gamma \left(\frac{2}{3}\right)^3 \Gamma \left(\frac{11}{6}\right)}+\frac{3\Gamma \left(\frac{1}{3}\right)
\Gamma \left(\frac{7}{6}\right)^2}{2 \sqrt{\pi }\Gamma \left(\frac{2}{3}\right)^2 \Gamma \left(\frac{5}{6}\right)}),
\nonumber \\
\mathcal{B}_{3}&=&\frac{1}{\mathcal{B}_{1}} \Bigg[ \frac{243\Gamma\qty(\frac{7}{6})^{7}}{2\pi^{5/2}\Gamma
\qty(\frac{2}{3})^{7}}\qty(\frac{3\Gamma\qty(\frac{2}{3})}{8\Gamma\qty(\frac{7}{6})}-\frac{\Gamma\qty(\frac{1}{3})
\Gamma\qty(\frac{4}{3})}{\Gamma\qty(\frac{5}{6})\Gamma\qty(\frac{11}{6})}) +    \nonumber  \\
&& \frac{\pi ^{3/2}\Gamma \left(-\frac{1}{3}\right) \Gamma \left(\frac{2}{3}\right)^2 \left(\frac{39366 \Gamma
\left(\frac{7}{6}\right)^{10} \Gamma \left(\frac{4}{3}\right)^2}{\pi^4 \Gamma \left(\frac{2}{3}\right)^{10} \Gamma
\left(\frac{11}{6}\right)^2}-\frac{26244 \Gamma \left(\frac{7}{6}\right)^9 \left(\sqrt{\pi } \Gamma \left(\frac{7}{6}\right)
\Gamma \left(\frac{4}{3}\right)^2-2 \Gamma \left(\frac{2}{3}\right) \Gamma \left(\frac{11}{6}\right)^2\right)}{\pi ^{9/2}
\Gamma \left(\frac{2}{3}\right)^{10} \Gamma \left(\frac{11}{6}\right)^2}\right)}{216 \Gamma \left(\frac{1}{6}\right)
\Gamma \left(\frac{7}{6}\right)^2} \Bigg].
\end{eqnarray}

\noindent Finally, the coefficients in (\ref{HEED4::LT}) [D$4$-brane] are written as
\begin{align}\label{app::D4}
\begin{split}
\mathcal{E}_{1}&=\frac{256 \pi ^{5/2}\Gamma \left(\frac{-2}{5}\right) \Gamma \left(\frac{3}{5}\right)^4}
{3125 \Gamma \left(\frac{1}{10}\right) \Gamma \left(\frac{11}{10}\right)^4} , \\
\mathcal{E}_{2}&=\frac{1}{\mathcal{E}_{1}} \frac{5\Gamma\qty(\frac{11}{10})^{3}}{8\sqrt{\pi}\Gamma
\qty(\frac{3}{5})^{3}}\qty(\frac{\Gamma\qty(\frac{-2}{5})\Gamma\qty(\frac{6}{5})}{\Gamma\qty(\frac{1}{10})
\Gamma\qty(\frac{17}{10})}+\frac{\Gamma\qty(\frac{1}{5})\Gamma\qty(\frac{3}{5})}{4\Gamma\qty(\frac{7}{10})
\Gamma\qty(\frac{11}{10})}), \\
\mathcal{E}_{3}&=\frac{1}{\mathcal{E}_{1}} \Bigg[ -\frac{78125 \Gamma \left(\frac{1}{5}\right) \Gamma
\left(\frac{11}{10}\right)^9 \Gamma \left(\frac{6}{5}\right)}{131072 \pi ^{7/2}\Gamma\left(\frac{3}{5}\right)^9
\Gamma \left(\frac{7}{10}\right) \Gamma \left(\frac{17}{10}\right)} + \frac{234375\Gamma \left(\frac{4}{5}\right)
\Gamma \left(\frac{11}{10}\right)^8}{524288 \pi ^{7/2}\Gamma \left(\frac{3}{5}\right)^8 \Gamma\left(\frac{13}{5}
\right)} +    \\
& \frac{64 \pi ^{5/2}\Gamma \left(-\frac{2}{5}\right) \Gamma \left(\frac{3}{5}\right)^4 \left(\frac{3662109375
\Gamma \left(\frac{6}{5}\right)^2 \Gamma \left(\frac{11}{10}\right)^{14}}{8388608 \pi^6 \Gamma
\left(\frac{3}{5}\right)^{14} \Gamma \left(\frac{17}{10}\right)^2}+\frac{732421875 \left(3 \Gamma \left(\frac{3}{5}\right)
\Gamma \left(\frac{17}{10}\right)^2 \Gamma \left(\frac{9}{5}\right)-2 \Gamma \left(\frac{11}{10}\right) \Gamma
\left(\frac{6}{5}\right)^2 \Gamma \left(\frac{23}{10}\right)\right) \Gamma \left(\frac{11}{10}\right)^{13}}{8388608 \pi^6
\Gamma \left(\frac{3}{5}\right)^{14} \Gamma \left(\frac{17}{10}\right)^2 \Gamma \left(\frac{23}{10}\right)}
\right)}{9375 \Gamma \left(\frac{1}{10}\right) \Gamma \left(\frac{11}{10}\right)^4} .
\end{split}
\end{align}

The coefficients $\text{a}_{i}$ in (\ref{eqn::cric}) can be expressed in terms of the constants appearing in
(\ref{app::D1})-(\ref{app::D4}) as follows:

\begin{table}[h]\label{Tab::A}
\centering
\begin{tabular}{ | l | l | l | l | l | l | } 
\hline
Background & $\text{a}_{0}$ & $\text{a}_{1}$ & $\text{a}_{2}$ & $\text{a}_{3}$ & $\text{a}_{4}$ \\ 
\hline \hline
D$1$ & $\mathcal{C}_{1}$ & $-\Theta_{p=1}/2$ & $2$ & $\mathcal{C}_{1}\mathcal{C}_{2}$ &
$\mathcal{C}_{1}\mathcal{C}_{3}$  \\ [8pt]
\hline
D$2$ & $\mathcal{D}_{1}$ & $-\Theta_{p=2}/2$ & $9/8$ & $\mathcal{D}_{1}\mathcal{D}_{2}$ &
$\mathcal{D}_{1}\mathcal{D}_{3}$    \\ [8pt]
\hline
D$3$ & $\mathcal{B}_{1}$ & $-\Theta_{p=3}/2$ & $1/2$ & $\mathcal{B}_{1}\mathcal{B}_{2}$ &
$\mathcal{B}_{1}\mathcal{B}_{3}$  \\ [8pt]
\hline
D$4$ & $\mathcal{E}_{1}$ & $-\Theta_{p=4}/2$ & $1/8$ & $\mathcal{E}_{1}\mathcal{E}_{2}$ & 
$\mathcal{E}_{1}\mathcal{E}_{3}$  \\ [8pt] \hline
\end{tabular}
\caption{Values of the coefficients $\text{a}_{i}$ in (\ref{eqn::cric})} corresponding to different
background configurations. Here $\Theta_{p=\#}$ corresponds to the value of the coefficient
$\Theta$ in (\ref{HEE::HT}) for different values of $p(=1,2,3,4)$.
\end{table}




\begin{thebibliography}{99}\label{biblio}

\bibitem{Maldacena:1997re}
J.~M.~Maldacena,
``The Large N limit of superconformal field theories and supergravity,''
Int. J. Theor. Phys. \textbf{38} (1999), 1113-1133, Adv. Theor. Math. Phys.
2 (1998) 231-252,
doi:10.1023/A:1026654312961
[arXiv:hep-th/9711200 [hep-th]].

\bibitem{Witten:1998qj}
E.~Witten,
``Anti-de Sitter space and holography,''
Adv. Theor. Math. Phys. \textbf{2} (1998), 253-291
doi:10.4310/ATMP.1998.v2.n2.a2
[arXiv:hep-th/9802150 [hep-th]].

\bibitem{Aharony:1999ti}
O.~Aharony, S.~S.~Gubser, J.~M.~Maldacena, H.~Ooguri and Y.~Oz,
``Large N field theories, string theory and gravity,''
Phys. Rept. \textbf{323} (2000), 183-386
doi:10.1016/S0370-1573(99)00083-6
[arXiv:hep-th/9905111 [hep-th]].

\bibitem{Ryu:2006bv}
S.~Ryu and T.~Takayanagi,
``Holographic derivation of entanglement entropy from AdS/CFT,''
Phys. Rev. Lett. \textbf{96} (2006), 181602
doi:10.1103/PhysRevLett.96.181602
[arXiv:hep-th/0603001 [hep-th]].

\bibitem{Ryu:2006ef}
S.~Ryu and T.~Takayanagi,
``Aspects of Holographic Entanglement Entropy,''
JHEP \textbf{08} (2006), 045
doi:10.1088/1126-6708/2006/08/045
[arXiv:hep-th/0605073 [hep-th]].

\bibitem{Hubeny:2007xt}
V.~E.~Hubeny, M.~Rangamani and T.~Takayanagi,
``A Covariant holographic entanglement entropy proposal,''
JHEP \textbf{07} (2007), 062
doi:10.1088/1126-6708/2007/07/062
[arXiv:0705.0016 [hep-th]].

\bibitem{Casini:2011kv}
H.~Casini, M.~Huerta and R.~C.~Myers,
``Towards a derivation of holographic entanglement entropy,''
JHEP \textbf{05} (2011), 036
doi:10.1007/JHEP05(2011)036
[arXiv:1102.0440 [hep-th]].

\bibitem{Faulkner:2013ana}
T.~Faulkner, A.~Lewkowycz and J.~Maldacena,
``Quantum corrections to holographic entanglement entropy,''
JHEP \textbf{11} (2013), 074
doi:10.1007/JHEP11(2013)074
[arXiv:1307.2892 [hep-th]].

\bibitem{Ben-Ami:2014gsa}
O.~Ben-Ami, D.~Carmi and J.~Sonnenschein,
``Holographic Entanglement Entropy of Multiple Strips,''
JHEP \textbf{11} (2014), 144
doi:10.1007/JHEP11(2014)144
[arXiv:1409.6305 [hep-th]].



\bibitem{Wolf:2007tdq}
M.~M.~Wolf, F.~Verstraete, M.~B.~Hastings and J.~I.~Cirac,
``Area Laws in Quantum Systems: Mutual Information and Correlations,''
Phys. Rev. Lett. \textbf{100} (2008) no.7, 070502
doi:10.1103/PhysRevLett.100.070502
[arXiv:0704.3906 [quant-ph]].

\bibitem{Headrick:2010zt}
M.~Headrick,
``Entanglement Renyi entropies in holographic theories,''
Phys. Rev. D \textbf{82} (2010), 126010
doi:10.1103/PhysRevD.82.126010
[arXiv:1006.0047 [hep-th]].

\bibitem{Fischler:2012uv}
W.~Fischler, A.~Kundu and S.~Kundu,
``Holographic Mutual Information at Finite Temperature,''
Phys. Rev. D \textbf{87} (2013) no.12, 126012
doi:10.1103/PhysRevD.87.126012
[arXiv:1212.4764 [hep-th]].



\bibitem{Takayanagi:2017knl}
T.~Takayanagi and K.~Umemoto,
``Entanglement of purification through holographic duality,''
Nature Phys. \textbf{14} (2018) no.6, 573-577
doi:10.1038/s41567-018-0075-2
[arXiv:1708.09393 [hep-th]].

\bibitem{Nguyen:2017yqw}
P.~Nguyen, T.~Devakul, M.~G.~Halbasch, M.~P.~Zaletel and B.~Swingle,
``Entanglement of purification: from spin chains to holography,''
JHEP \textbf{01} (2018), 098
doi:10.1007/JHEP01(2018)098
[arXiv:1709.07424 [hep-th]].

\bibitem{Terhal:2002bm} B.~M.~Terhal, M.~Horodecki, D.~W.~Leung and D.~P.~
DiVincenzo, ``The entanglement of purification," J.~Math.~Phys.~ \textbf{43}
(2002) 4286 [quan-ph/0202044].

\bibitem{Kudler-Flam:2018qjo}
J.~Kudler-Flam and S.~Ryu,
``Entanglement negativity and minimal entanglement wedge cross sections in
holographic theories,''
Phys. Rev. D \textbf{99} (2019) no.10, 106014
doi:10.1103/PhysRevD.99.106014
[arXiv:1808.00446 [hep-th]].

\bibitem{Dutta:2019gen}
S.~Dutta and T.~Faulkner,
``A canonical purification for the entanglement wedge cross-section,''
[arXiv:1905.00577 [hep-th]].

\bibitem{Jeong:2019xdr}
H.~S.~Jeong, K.~Y.~Kim and M.~Nishida,
``Reflected Entropy and Entanglement Wedge Cross Section with the First Order
Correction,''
JHEP \textbf{12} (2019), 170
doi:10.1007/JHEP12(2019)170
[arXiv:1909.02806 [hep-th]].

\bibitem{BabaeiVelni:2019pkw}
K.~Babaei Velni, M.~R.~Mohammadi Mozaffar and M.~H.~Vahidinia,
``Some Aspects of Entanglement Wedge Cross-Section,''
JHEP \textbf{05} (2019), 200
doi:10.1007/JHEP05(2019)200
[arXiv:1903.08490 [hep-th]].

\bibitem{Jokela:2019ebz}
N.~Jokela and A.~P\"{o}nni,
``Notes on entanglement wedge cross sections,''
JHEP \textbf{07} (2019), 087
doi:10.1007/JHEP07(2019)087
[arXiv:1904.09582 [hep-th]].

\bibitem{Chakrabortty:2020ptb}
S.~Chakrabortty, S.~Pant and K.~Sil,
``Effect of back reaction on entanglement and subregion volume complexity in
strongly coupled plasma,''
JHEP \textbf{06} (2020), 061
doi:10.1007/JHEP06(2020)061
[arXiv:2004.06991 [hep-th]].

\bibitem{Huang:2019zph}
Y.~f.~Huang, Z.~j.~Shi, C.~Niu, C.~y.~Zhang and P.~Liu,
``Mixed State Entanglement for Holographic Axion Model,''
Eur. Phys. J. C \textbf{80} (2020) no.5, 426
doi:10.1140/epjc/s10052-020-7921-y
[arXiv:1911.10977 [hep-th]].

\bibitem{Boruch:2020wbe}
J.~Boruch,
``Entanglement wedge cross-section in shock wave geometries,''
JHEP \textbf{07} (2020), 208
doi:10.1007/JHEP07(2020)208
[arXiv:2006.10625 [hep-th]].

\bibitem{Fu:2020oep}
G.~Fu, P.~Liu, H.~Gong, X.~M.~Kuang and J.~P.~Wu,
``Informational properties for Einstein-Maxwell-Dilaton Gravity,''
[arXiv:2007.06001 [hep-th]].

\bibitem{Yang:2018gfq}
R.~Q.~Yang, C.~Y.~Zhang and W.~M.~Li,
``Holographic entanglement of purification for thermofield double states
and thermal quench,''
JHEP \textbf{01} (2019), 114
doi:10.1007/JHEP01(2019)114
[arXiv:1810.00420 [hep-th]].

\bibitem{Kudler-Flam:2020url}
J.~Kudler-Flam, Y.~Kusuki and S.~Ryu,
``Correlation measures and the entanglement wedge cross-section after
quantum quenches in two-dimensional conformal field theories,''
JHEP \textbf{04} (2020), 074
doi:10.1007/JHEP04(2020)074
[arXiv:2001.05501 [hep-th]].

\bibitem{BabaeiVelni:2020wfl}
K.~Babaei Velni, M.~R.~Mohammadi Mozaffar and M.~H.~Vahidinia,
``Evolution of entanglement wedge cross section following a global quench,''
JHEP \textbf{08} (2020), 129
doi:10.1007/JHEP08(2020)129
[arXiv:2005.05673 [hep-th]].


\bibitem{Itzhaki:1998dd}
N.~Itzhaki, J.~M.~Maldacena, J.~Sonnenschein and S.~Yankielowicz,
``Supergravity and the large N limit of theories with sixteen supercharges,''
Phys. Rev. D \textbf{58} (1998), 046004
doi:10.1103/PhysRevD.58.046004
[arXiv:hep-th/9802042 [hep-th]].

\bibitem{Boonstra:1998mp}
H.~J.~Boonstra, K.~Skenderis and P.~K.~Townsend,
``The domain wall / QFT correspondence,''
JHEP \textbf{01} (1999), 003
doi:10.1088/1126-6708/1999/01/003
[arXiv:hep-th/9807137 [hep-th]].

K.~Skenderis,
``Field theory limit of branes and gauged supergravities,''
Fortsch. Phys. \textbf{48} (2000), 205-208
doi:10.1002/(SICI)1521-3978(20001)48:1/3<205::AID-PROP205>3.0.CO;2-F
[arXiv:hep-th/9903003 [hep-th]].

\bibitem{Kanitscheider:2008kd}
I.~Kanitscheider, K.~Skenderis and M.~Taylor,
``Precision holography for non-conformal branes,''
JHEP \textbf{09} (2008), 094
doi:10.1088/1126-6708/2008/09/094
[arXiv:0807.3324 [hep-th]].

\bibitem{Pang:2013lpa}
D.~W.~Pang,
``Entanglement thermodynamics for nonconformal D-branes,''
Phys. Rev. D \textbf{88} (2013) no.12, 126001
doi:10.1103/PhysRevD.88.126001
[arXiv:1310.3676 [hep-th]].

\bibitem{Pang:2015lka}
D.~W.~Pang,
``Corner contributions to holographic entanglement entropy in non-conformal
backgrounds,''
JHEP \textbf{09} (2015), 133
doi:10.1007/JHEP09(2015)133
[arXiv:1506.07979 [hep-th]].

\bibitem{Narayan:2012hk}
K.~Narayan,
``On Lifshitz scaling and hyperscaling violation in string theory,''
Phys. Rev. D \textbf{85} (2012), 106006
doi:10.1103/PhysRevD.85.106006
[arXiv:1202.5935 [hep-th]].

\bibitem{Singh:2012un}
H.~Singh,
``Lifshitz/Schr\"{o}dinger Dp-branes and dynamical exponents,''
JHEP \textbf{07} (2012), 082
doi:10.1007/JHEP07(2012)082
[arXiv:1202.6533 [hep-th]].

\bibitem{Narayan:2013qga}
K.~Narayan,
``Non-conformal brane plane waves and entanglement entropy,''
Phys. Lett. B \textbf{726} (2013), 370-374
doi:10.1016/j.physletb.2013.07.061
[arXiv:1304.6697 [hep-th]].


\bibitem{Mishra:2016yor}
R.~Mishra and H.~Singh,
``Entanglement asymmetry for boosted black branes and the bound,''
Int. J. Mod. Phys. A \textbf{32} (2017) no.16, 1750091
doi:10.1142/S0217751X17500919
[arXiv:1603.06058 [hep-th]].

\bibitem{vanNiekerk:2011yi}
A.~van Niekerk,
``Entanglement Entropy in NonConformal Holographic Theories,''
[arXiv:1108.2294 [hep-th]].

\bibitem{Fischler:2012ca}
W.~Fischler and S.~Kundu,
``Strongly Coupled Gauge Theories: High and Low Temperature Behavior of
Non-local Observables,''
JHEP \textbf{05} (2013), 098
doi:10.1007/JHEP05(2013)098
[arXiv:1212.2643 [hep-th]].

\bibitem{Kundu:2016dyk}
S.~Kundu and J.~F.~Pedraza,
``Aspects of Holographic Entanglement at Finite Temperature and Chemical Potential,''
JHEP \textbf{08} (2016), 177
doi:10.1007/JHEP08(2016)177
[arXiv:1602.07353 [hep-th]].

\bibitem{Wald:1984}
Robert~M.~Wald,
General relativity, Chicago Univ. Press, Chicago, IL, 1984.










\end{thebibliography}
\end{document}